\documentclass[aps,prb,twocolumn,superscriptaddress,showpacs]{revtex4-2}
\usepackage[colorlinks,linkcolor=blue,citecolor=blue,urlcolor=blue,breaklinks=true]{hyperref}
\usepackage{amsmath,amssymb,graphics,epsfig,epstopdf,color,verbatim,braket,tabularx,breakurl}
\usepackage{multirow}
\usepackage{diagbox}
\usepackage{subfig}
\usepackage[lined,boxed,commentsnumbered]{algorithm2e}
\usepackage{stmaryrd}
\usepackage[normalem]{ulem}
\usepackage{float}
\usepackage{ragged2e}
\usepackage{placeins}
\usepackage{babel}
\newcommand{\spins}{\boldsymbol{S}}

\begin{document}
\title{Single-layer framework of variational tensor network states }

\author{Hongyu Chen}
\address{School of Physics, Renmin University of China, Beijing 100872, China}

\author{Yangfeng Fu}
\address{School of Physics, Renmin University of China, Beijing 100872, China}

\author{Weiqiang Yu}
\address{School of Physics, Renmin University of China, Beijing 100872, China}
\address{Key Laboratory of Quantum State Construction and Manipulation (Ministry of Education), Renmin University of China, Beijing, 100872, China}

\author{Rong Yu}
\address{School of Physics, Renmin University of China, Beijing 100872, 
China}
\address{Key Laboratory of Quantum State Construction and Manipulation (Ministry 
of Education), Renmin University of China, Beijing, 100872, China}

\author{Z. Y. Xie}
\email{qingtaoxie@ruc.edu.cn}
\address{School of Physics, Renmin University of China, Beijing 100872, China}
\address{Key Laboratory of Quantum State Construction and Manipulation (Ministry 
of Education), Renmin University of China, Beijing, 100872, China}

\begin{abstract}      
      We propose a single-layer tensor network framework 
      for the variational determination of ground states in two-dimensional 
      quantum lattice models. 
      By combining the nested tensor network method [Phys. Rev. B \textbf{96}, 
      045128 (2017)] 
      with the automatic differentiation technique, our 
      approach can reduce the 
      computational cost by three orders of magnitude in 
      bond dimension, and 
      therefore 
      enables highly efficient variational ground-state calculations. 
      We demonstrate 
      the capability of this framework through 
      two quantum spin models: the antiferromagnetic Heisenberg model on a 
      square lattice and the frustrated Shastry-Sutherland model. Even without 
      GPU 
      acceleration or symmetry 
      implementation, we have achieved 
      a bond dimension 
      of nine and 
      obtained accurate ground-state energy and consistent order parameters 
      compared to 
      prior studies. In particular, we confirm  
      the existence of an intermediate empty-plaquette valence bond solid 
      ground state 
      in the Shastry-Sutherland model. We have further discussed the 
      convergence of the algorithm 
      and 
      its potential 
      improvements. 
      Our work provides 
      a promising route 
      for large-scale tensor network calculations of two-dimensional quantum 
      systems.
       		
\end{abstract}
\maketitle
	
\section{Introduction} 
\label{Sec:Intro}
Two-dimensional (2D) strongly correlated systems, such as interacting fermionic 
systems and 
frustrated 
quantum magnets, pose significant challenges for theoretical studies. 
Consequently, developing efficient and accurate numerical methods for these 
models 
constitutes a crucial branch of theoretical condensed matter physics. 
In recent decades, tensor network methods \cite{XiangBook, SimonBook} 
have attracted increasing attention. By combining the tensor-network ansatz 
\cite{PEPS2004, OrusNatRev} stemming from the quantum entanglement, 
the contraction strategy 
originating from 
renormalization group \cite{Orus2014, RanBook}, and 
modern optimization techniques developed in 
artificial intelligence
\cite{OptBook}, 
tensor network methods have been successfully applied to diverse phenomena, 
including
quantum 
magnetism \cite{KagomePRL2017, J1J2-LWY, NingXi-SSM2023, LY-PRB2025}, 
superconductivity \cite{TNStJ, TNSHubbard, Corboz-MulCTMRG}, topological order 
\cite{TNStopo, CSL}, and quantum field theory \cite{CMPS, CTNS}. 

For studying 
ground states of 
2D quantum systems using tensor networks, 
three strategies are commonly adopted.
The first 
maps the 
2D quantum model to a three-dimensional classical one 
\cite{QCmap}, 
converting the quantum expectation value calculation into a statistical physics  
problem \cite{HOTRG}. The second 
employs 
imaginary-time evolution 
with wave function update schemes 
\cite{SU, CU, FU, FFU} to 
obtain 
the ground state $|\Psi_g\rangle$ in the tensor-network representation, and then 
evaluate the expectation value $\langle\hat{O}\rangle$ 
of the operator $\hat{O}$ by contracting the related tensor networks. The 
 third one is the 
straightforward variational minimization: 
Starting from 
an arbitrary tensor network state 
$|\Psi\rangle$, the ground state of the system $|\Psi_g\rangle$ and its energy 
$E_g$ are obtained by minimizing the
energy estimator 
\begin{eqnarray}
E=\frac{\langle\Psi|\hat{H}|\Psi\rangle}{\langle\Psi|\Psi\rangle}   
\label{Eq:ExpH}
\end{eqnarray}
with respect to tensor parameters 
\cite{PEPS2004}. 
Recently, this approach has become increasingly popular because of the advent of the 
automatic differentiation technique \cite{ADreview}, 
which, as a mature and widely applied algorithm \cite{NeuralBook}, can 
efficiently calculate the gradient $\frac{dE}{d|\Psi\rangle}$ \cite{LW-PTNS2019} 
and easily integrate with gradient-based optimization method \cite{OptBook} via 
various open-source 
packages \cite{Pytorch, Zygote}.

A central challenge in all three approaches
is the contraction of 
tensor networks. Unfortunately, contracting a general tensor network in more 
than one dimension is a $\#P$-complete hard problem \cite{PEPSComplex} that 
cannot be performed exactly for an infinite system even with translational 
invariance. To this end, a series of approximation algorithms based on the 
renormalization group idea have been developed \cite{RanBook, XiangBook}, among 
which the corner transfer-matrix renormalization group (CTMRG) 
\cite{Nishino-CTMRG, Vidal-CTM, Corboz-MulCTMRG, Vari-CTMRG} 
has been widely used.  In two dimensions, the cost of the CTMRG 
algorithm 
scales as 
$D^{10}$ in memory 
and $D^{12}$ in computation with the bond dimension $D$ when 
no symmetry or partial matrix decomposition is employed. 
The cost is 
extremely high for variational calculations, which 
typically require hundreds of rounds of such contractions. The problem becomes 
even more serious for minimization with the automatic differentiation since it 
needs huge memory to store 
intermediate data 
for the gradient ($\frac{dE}{d|\Psi\rangle}$) calculation. 
This issue pose severe limitation for the bond dimension $D$ in variational 
minimization 
employing automatic differentiation. 

To overcome this limitation, 
here we 
propose a single-layer framework of variational tensor network state 
calculation that incorporates the  nested tensor network (NTN) method 
\cite{NTN2017}. By 
reformulating the double-layer structures of numerator and denominator in 
Eq.~(\ref{Eq:ExpH}) 
into nested single-layer structures, 
the NTN method reduces the memory and computation cost of tensor network 
contraction by two and three orders of $D$, respectively. This allows us to 
perform variational minimization with significantly larger bond dimension. 
Our proposal works as follows: For any given 
state $|\Psi\rangle$, we utilize the NTN to calculate the energy defined in 
Eq.~(\ref{Eq:ExpH}), then calculate the gradient $\frac{dE}{d|\Psi\rangle}$, 
and finally update the wave function $|\Psi\rangle$ using the gradient 
information as usual \cite{LW-PTNS2019}. This procedure can be iterated many 
times until the calculated energy converges. 

We 
validate this method on two benchmark models defined on infinite 2D lattices:
the $S=1/2$ square-lattice antiferromagnetic Heisenberg model and 
the frustrated Shastry-Sutherland model. The former has a long-range antiferromagnetically 
ordered ground state \cite{HM-MC}, while the latter 
hosts a 
series of ground states when tuning the strength of spin frustration 
$J/J^\prime$ \cite{Corboz-SSM, NingXi-SSM2023}. In both cases, with the help of 
NTN, we can efficiently reach $D = 9$ even without any GPU utilization 
or symmetry 
implementation, which is difficult in previous variational calculations 
\cite{LW-PTNS2019, NingXi-SSM2023, LY-PRB2025}. The obtained ground-state energy 
and order parameters are consistent with previous calculations, but with better 
convergence. In particular, we 
provide clear evidence for an empty-plaquette valence 
bond solid (VBS) phase in the ground-state phase diagram of the 
Shastry-Sutherland model.

The rest of the paper is organized as follows. In Sec.~\ref{Sec:Method} 
we first briefly review the automatic differentiation-assisted variational 
calculation of tensor network state and the NTN method, 
then introduce our 
single-layer framework. 
In Sec.~\ref{Sec:Results}, we 
present the 
numerical results of the Heisenberg model and the Shastry-Sutherland model in 
detail. Finally, we summarize and conclude in Sec.~\ref{Sec:Outlook}.

\section{Method} 
\label{Sec:Method}
In this section, we first review the basics of automatic 
differentiation-assisted variational calculation of tensor network 
representation of the ground state \cite{LW-PTNS2019}, the underlying 
double-layer structure in Eq.~(\ref{Eq:ExpH}), and the NTN method 
\cite{NTN2017}, 
then introduce the single-layer framework by combining the two parts 
together. 

\subsection{Variational calculation assisted by automatic differentiation}
\label{Sec:AD}
\begin{figure}[!h]
\centering
\includegraphics[scale=0.35]{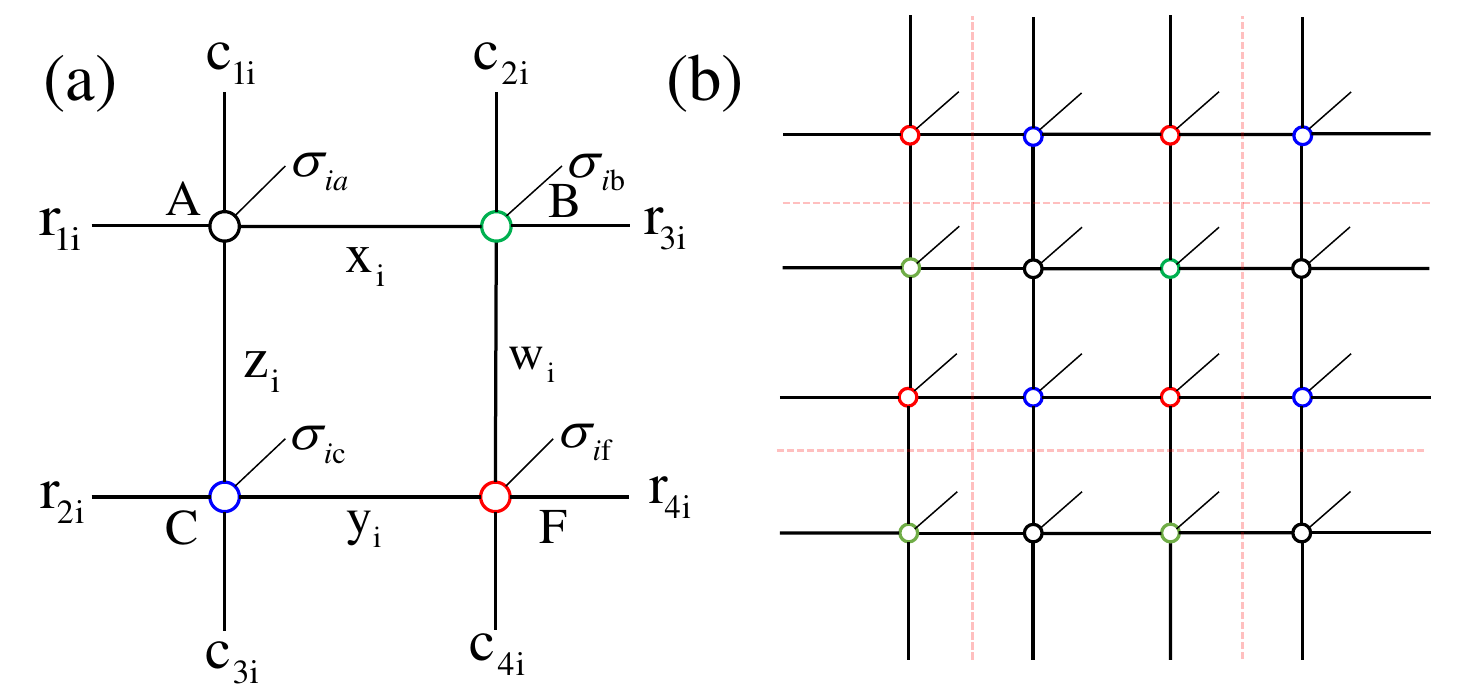}
\caption{An illustration of the infinite PEPS ansatz with a $2\times 2$ 
sublattice, as expressed in Eq.~(\ref{Eq:PEPS}). (a) A $2\times 2$ unit cell.  
(b) Infinite PEPS wave function with translational invariance. Local tensors 
defined on the dots with the same color are identical. Dashed lines separate 
different unit cells.}
\label{Fig:PEPS}
\end{figure}

For concreteness, we take the projected entangled pair state (PEPS) on a square 
lattice as an example to explain how to determine the ground state of a quantum 
spin system in the PEPS representation. 
Without loss of generality, suppose we have an infinite system hosting a 
$2\times 2$ sublattice structure, then the infinite PEPS ansatz can be written 
as
\begin{eqnarray}
|\Psi\rangle = &&\sum_{\{\sigma\}} \Big[\mathrm{Tr}\prod_{i}A^{(i)}_{r_{1i}x_ic_{1i}z_i}[\sigma_{ia}]B^{(i)}_{x_ir_{3i}c_{2i}w_i}[\sigma_{ib}]\nonumber \\
&&C^{(i)}_{r_{2i}y_iz_ic_{3i}}[\sigma_{ic}]
F^{(i)}_{y_ir_{4i}w_ic_{4i}}[\sigma_{if}]\Big]\Big|~...\sigma_{ia}\sigma_{ib}\sigma_{ic}\sigma_{if}...\Big\rangle \nonumber \\
\label{Eq:PEPS}
\end{eqnarray}
where the superscript $i$ denotes the unit cell index, and $A^{(i)}$, 
$B^{(i)}$, $C^{(i)}$, $F^{(i)}$ are the four distinct tensors defined in the 
$i$-th unit cell. 
$\textrm{Tr}$ means 
summing over all the tensor subscripts 
that correspond to the link indices, and 
$\sum$ means summation over all the physical spin configurations $\{\sigma\}$. 
A sketch of the $|\Psi\rangle$ is illustrated in Fig.~\ref{Fig:PEPS}.

According to Eq.~(\ref{Eq:PEPS}), any quantum state $|\Psi\rangle$ satisfying 
the area law 
can be compatibly expressed as a superposition of spin configurations, where 
the coefficient of each term is represented as a tensor-network summation 
denoted by $\textrm{Tr}$. For an infinite system with translational symmetry, 
we assume the local tensors in different unit cells are the same, as 
illustrated in Fig.~\ref{Fig:PEPS}(b). Therefore, for any given local tensors, 
we can introduce a shorthand
\begin{eqnarray}
X = [\vec{A}, \vec{B}, \vec{C}, \vec{F}]
\end{eqnarray}
where $\vec{A}$ is the local tensor $A$ reshaped as a vector residing in the 
$(D^4d)$-dimensional space, and so are $\vec{B}$, $\vec{C}$ and $\vec{F}$. 
X is then a $(4D^4d)$-dimensional vector storing all the variational parameters 
necessary to specify a quantum state $|\Psi(X)\rangle$ represented as 
Eq.~(\ref{Eq:PEPS}). Here $D$ is the 
bond dimension, which is the upper bound of all the subscripts appeared in 
Eq.~(\ref{Eq:PEPS}), and $d$ is the physical dimension for each $\sigma$. A 
tensor with a larger $D$ 
contains more parameters, and is expected to have stronger representation power 
and better accuracy for a state. However, 
this also means heavier cost (discussed in Sec.~\ref{Sec:NTN}). To find the 
PEPS representation of the ground state $|\Psi_g\rangle$ of a Hamiltonian 
system, we need to solve the following optimization problem
\begin{eqnarray}
\mathrm{arg~min}_{X} E(X)
\label{Eq:OptPro}
\end{eqnarray}
where $E(X)$ is determined by Eq.~(\ref{Eq:ExpH}) but now 
its dependence of $X$ is explicitly written.

In this work, we follow the 
variational minimization approach outlined 
in Sec.~\ref{Sec:Intro}. 
Starting from an initial $|\Psi(X_0)\rangle$, 
we can use well-established techniques, such as the CTMRG \cite{Nishino-CTMRG, 
Vidal-CTM, Corboz-MulCTMRG, Vari-CTMRG}, to compute the energy $E(X)$. In this 
procedure, 
when the automatic differentiation technique is used. Then, as long as $E$ is 
obtained successfully, 
we can calculate the gradient $\frac{\partial{E(X)}}{\partial{X}}\big|_{X=X_0}$ 
conveniently by simply calling some open source packages, such as PyTorch 
\cite{Pytorch} and Zygote \cite{Zygote}, without any extra operation. After 
the gradient information is obtained, we can update $X$ using classic 
optimization algorithms, such as L-BFGS \cite{OptBook}, 
to get a better 
trial $X_1$ and the corresponding wave function $|\Psi(X_1)\rangle$. Similary, 
we can evaluate $E(X_1)$, calculate 
$\frac{\partial{E(X)}}{\partial{X}}\big|_{X=X_1}$, and 
then obtain a better 
$X_2$ and $|\Psi(X_2)\rangle$. This iteration can be repeated as many as we 
want 
until the gradient $\frac{\partial{E(X)}}{\partial{X}}$ is sufficiently small 
or the energy $E$ converges within some tolerance. The converged $X^*$ 
yields an approximate solution of Eq.~(\ref{Eq:OptPro}) and 
thereby provides an approximate 
representation of $|\Psi_g\rangle$.  

In essence, the reverse mode of automatic differentiation \cite{ADreview} uses the chain rule of derivative, usually referred to as backpropagation \cite{BP1986} in optimization engineering, to compute the gradient information. Formally, in this case
\begin{eqnarray}
\frac{\partial{E}}{\partial{X}} = \frac{\partial{E}}{\partial{T^{(n)}}}\frac{\partial{T^{(n)}}}{\partial{T^{(n-1)}}}\frac{\partial{T^{(n-1)}}}{\partial{T^{(n-2)}}}...\frac{\partial{T^{(2)}}}{\partial{T^{(1)}}}\frac{\partial{T^{(1)}}}{\partial{X}}
\label{Eq:AD}
\end{eqnarray}
where the $\{T^{(i)},i=1,2,...n\}$ are the intermediate variables that are 
involved in the process of evaluating $E$, e.g., the temporary local tensors 
generated in the CTMRG iterations. We assume that $n$ related local tensors in 
Eq.~(\ref{Eq:AD}) have been generated. In order to 
compute the 
gradient $\frac{\partial{E}}{\partial{X}}$, 
these intermediate tensors generally need to 
be stored in memory. In the 
tensor network framework, if 
the transformations involved in the energy-evaluation contraction algorithms 
are treated as independent parameters, 
automatic differentiation is then equivalent to the backward iteration 
procedure of the second renormalization group method \cite{SRGAD2020} for 
obtaining the global environment. Therefore, in this sense, this kind of 
variational tensor network approach tries to solve Eq.~(\ref{Eq:OptPro}) by 
global optimization, and that is one of the reasons why this method can be 
successfully applied to study a series of strongly correlated systems 
\cite{LW-PTNS2019}, e.g., quantum spin-orbital liquid \cite{LY-PRB2023}, 
altermagnetism \cite{LY-PRB2025}, deconfined quantum critical point 
\cite{NingXi-SSM2023}, even with a small bond dimension.
  
\subsection{Review of the NTN method}
\label{Sec:NTN}    
\begin{figure}[!h]
	\centering
	\includegraphics[scale=0.35]{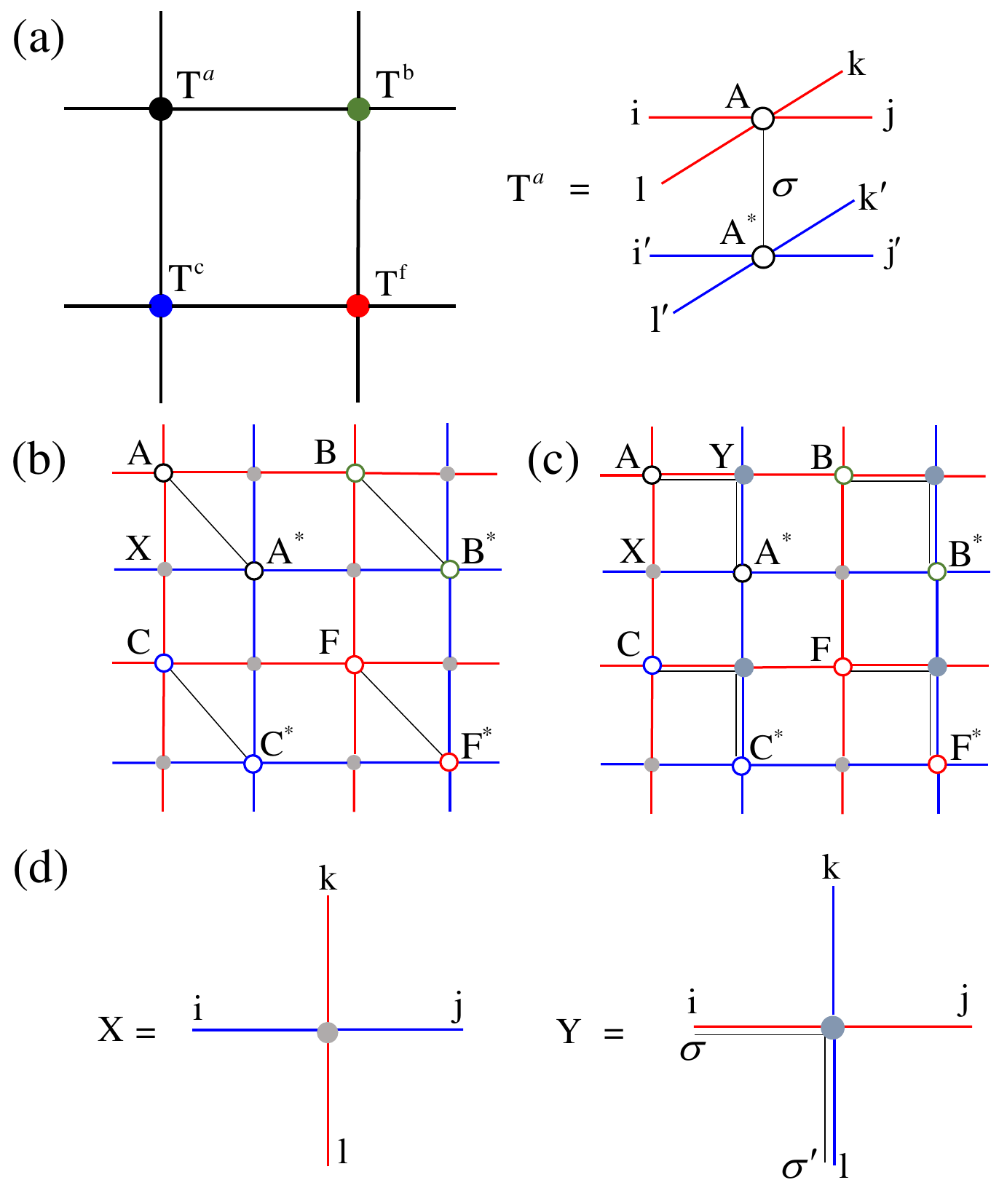}
	\caption{An illustration of the NTN method. (a) The unit cell of the reduced network $\langle\Psi|\Psi\rangle$, corresponding to the unit cell of $|\Psi\rangle$ in Fig.~\ref{Fig:PEPS}. The double-layer structure of $T$ is also shown, corresponding to Eq.~(\ref{Eq:Tred}). (b) The actual structure of (a), where $\langle\Psi|$ and $|\Psi\rangle$ are colored by blue and red, respectively. The slashed black lines denote physical indices $\{\sigma\}$. (c) The nested representation of (a) and (b) used in this work. Two additional tensors $X$ and $Y$ are introduced to form a compact single-layer tensor network. (d) The definition of $X$ and $Y$, as expressed in Eq.~(\ref{Eq:XY}). }
	\label{Fig:NTN}
\end{figure}

    As mentioned in the last section, the intermediate local tensors $\{T^{(i)}\}$ in Eq.~(\ref{Eq:AD}) have to be stored in memory before the gradient is obtained. This poses a great challenge for memory and strongly limits the bond dimension $D$ in the calculations. To see this, in Fig.~\ref{Fig:NTN}, we sketch $\langle\Psi|\Psi\rangle$ whose calculation cannot be avoided in evaluating Eq.~(\ref{Eq:ExpH}). Conventionally, to calculate $\langle\Psi|\Psi\rangle$, firstly one need to form a reduced tensor
\begin{eqnarray}
T^{a}_{ii',jj',kk',ll'}\equiv\sum_{\sigma}A_{i,j,k,l}[\sigma]A^*_{i',j',k',l'}[\sigma]
\label{Eq:Tred}
\end{eqnarray}
as illustrated in Fig.~\ref{Fig:NTN}(a). $T^{b}$, $T^{c}$, $T^{f}$ can be 
defined similarly corresponding to the $2\times 2$ unit cell in 
Fig.~\ref{Fig:PEPS}(a). Obviously, each local tensor $T$, and thus the resulting 
tensor network $\langle\Psi|\Psi\rangle$, has a \emph{double-layer} structure. 
That is, if the representation $|\Psi\rangle$ in Eq.~(\ref{Eq:PEPS}) has bond 
dimension $D$, then each index of $T^a$ in Eq.~(\ref{Eq:Tred}) and the resulting 
$\langle\Psi|\Psi\rangle$ have bond dimension $D^2$.  Therefore, evaluating the 
energy $E(X)$ requires contracting a tensor network with bond dimension $D^2$, 
which generally leads to 
a computational cost 
of about $D^{12}$ ~\cite{NTN2017} when the CTMRG algorithm is employed 
\cite{YJKao-CTMRG, DNSheng-CTMRG}. As explained in Sec.~\ref{Sec:AD}, the memory 
cost of storing all intermediate tensors generated in this procedure 
in automatic differentiation is 
typically 
$D^8$, which would expand fast with increasing $D$.  
Therefore, the substantial memory and computational demands impose a stringent 
constraint on the practically attainable bond dimension $D$ in automatic 
differentiation-based variational minimization.
    
    In order to solve this problem, the NTN method regards $\langle\Psi|\Psi\rangle$ as a \emph{single-layer} network with a nesting structure, as illustrated in Fig.~\ref{Fig:NTN}(c). To be specific, instead of summing over the physical index $\sigma$ and forming a reduced tensor $T^a$, as done in Eq.~(\ref{Eq:Tred}), NTN considers the physical indices $\{\sigma\}$ and the link indices equally by regarding $\{\sigma\}$ as special link indices connecting nearby conjugated tensors, e.g, $A^*$ in $\langle\Psi|$ and $A$ in $|\Psi\rangle$, as shown in Fig.~\ref{Fig:NTN}(b). In graphic terms, the bra state $\langle\Psi|$ (blue) and the ket state $|\Psi\rangle$ (red) lie in the same layer and are connected by $\{\sigma\}$. In order to avoid longer-range links, a virtual tensor $X$, equal to the direct product of two $\delta$ functions, is defined at each crossing point where the virtual links in $\langle\Psi|$ and $|\Psi\rangle$ intersect. Furthermore, to make the resulting network a square lattice, we modify one set of $X$ to $Y$ by combining an $X$ and a $\sigma$. Formally, as shown in Fig.~\ref{Fig:NTN}(d), $X$ and $Y$ can be defined as follows    
\begin{eqnarray}
X_{i,j,k,l} = \delta_{ij}\delta_{kl}, \quad Y_{i\sigma,j,k,l\sigma'} = \delta_{ij}\delta_{kl}\delta_{\sigma\sigma'}
\label{Eq:XY} 
\end{eqnarray}
where $\delta$ is the Kronecker-delta function. Finally, the nesting structure of $\langle\Psi|\Psi\rangle$ is illustrated in Fig.~\ref{Fig:NTN}(c) as a single-layer network. Any expectation value of a local operator, such as local terms in the numerator in Eq.~(\ref{Eq:ExpH}), i.e., $\langle\Psi|H_{ij}|\Psi\rangle$ where $H_{ij}$ is a bond Hamiltonian, can be represented similarly but with some impurity tensors.
    
    Though the idea is simple, the NTN method has a significant advantage that the bond dimension of $\langle\Psi|\Psi\rangle$ is no more than $Dd$, where $d$ is the physical dimension associated with each $\sigma$, instead of $D^2$ as in Fig.~\ref{Fig:NTN}(a).  Since the nesting structure is essentially a single-layer tensor network, widely used methods such as CTMRG can be employed to perform the contraction efficiently. For example, suppose the environment dimension used in CTMRG is of the same order as $D^2$ required by the reduced tensor network with bond dimension $D^2$. In that case, ignoring the scaling of $d$, the contraction of this nested single-layer tensor network has a memory cost of about $D^6$ and a computational cost of about $D^9$, which are much lower than those in the reduced method ($D^8$ and $D^{12}$) \cite{NTN2017}. 
    
    Bearing this advantage, the NTN method has been used in a series of studies to extend $D$ to larger values on the kagome lattice. For example, $D$ was extended to $25$ in Refs.~\cite{NTN2017, KagomePRL2017} to study the kagome spin liquid, also to $25$ in Ref.~\cite{YJKao-CTMRG} to study the kagome Heisenberg model with DM interactions, and to $15$ in Ref.~\cite{DNSheng-CTMRG} to study the kagome Heisenberg model with chiral interactions.  
     
\subsection{The single-layer framework of variational tensor network states}
\label{Sec:SLF}

    As far as we know, in previous studies, such as Refs.~\cite{NTN2017, 
    KagomePRL2017, YJKao-CTMRG, DNSheng-CTMRG} mentioned in Sec.~\ref{Sec:NTN}, 
    the NTN method was only utilized in either expectation value calculation or 
    updating the wave function in imaginary-time evolution, but was seldom 
    utilized in the variational calculation of the ground state wave function as 
    discussed in Sec.~\ref{Sec:AD}.
    
    In this work, we propose to incorporate the NTN method into the variational solution of the minimization problem, i.e., Eq.~(\ref{Eq:OptPro}). The overall framework is summarized in the following:
    
    (i). First, we need to prepare an initial PEPS $|\Psi(X_0)\rangle$ from which the optimization begins. Either an arbitrary state specified by a random $X_0$ (this work) or an approximate PEPS obtained by imaginary-time evolution \cite{SU, CU, FU, FFU} can be accepted. 
    
    (ii). Compute the energy $E(X_0)$ through Eq.~(\ref{Eq:ExpH}). where either boundary matrix product state \cite{Vidal-TEBD, Orus-TEBD} or CTMRG (this work) can be used to contract the single-layer tensor network $\langle\Psi(X_0)|\Psi(X_0)\rangle$ with nested structure explained in Sec.~\ref{Sec:NTN}. In this step, one can use automatic differentiation software packages, such as PyTorch \cite{Pytorch} and Zygote \cite{Zygote} (this work), to implement the contraction algorithm. Once $E(X_0)$ is obtained, the gradient $\frac{\partial E(X)}{\partial X}\big|_{X=X_0}$ can be evaluated through Eq.~(\ref{Eq:AD}) efficiently and automatically. 
    
    (iii). Using the gradient information obtained in step (ii), we can employ 
    gradient-based optimization algorithm, such as Adam and L-BFGS 
    \cite{OptBook} (this work), to get $X_1$, the updated estimate of the 
    solution of Eq.~(\ref{Eq:OptPro}). Then update the parameter by setting 
    $X_0=X_1$ and obtain a new PEPS $|\Psi(X_0)\rangle$.
    
    (iv). Repeat steps (ii) and (iii) until the energy $E$ converges or the gradient stops decreasing. Finally, we regard the output $X_1$ as an approximate solution of Eq.~(\ref{Eq:OptPro}) and $|\Psi(X_1)\rangle$ as an approximate ground state correspondingly.
    
\begin{figure}[!h]
	\centering
	\includegraphics[scale=0.35]{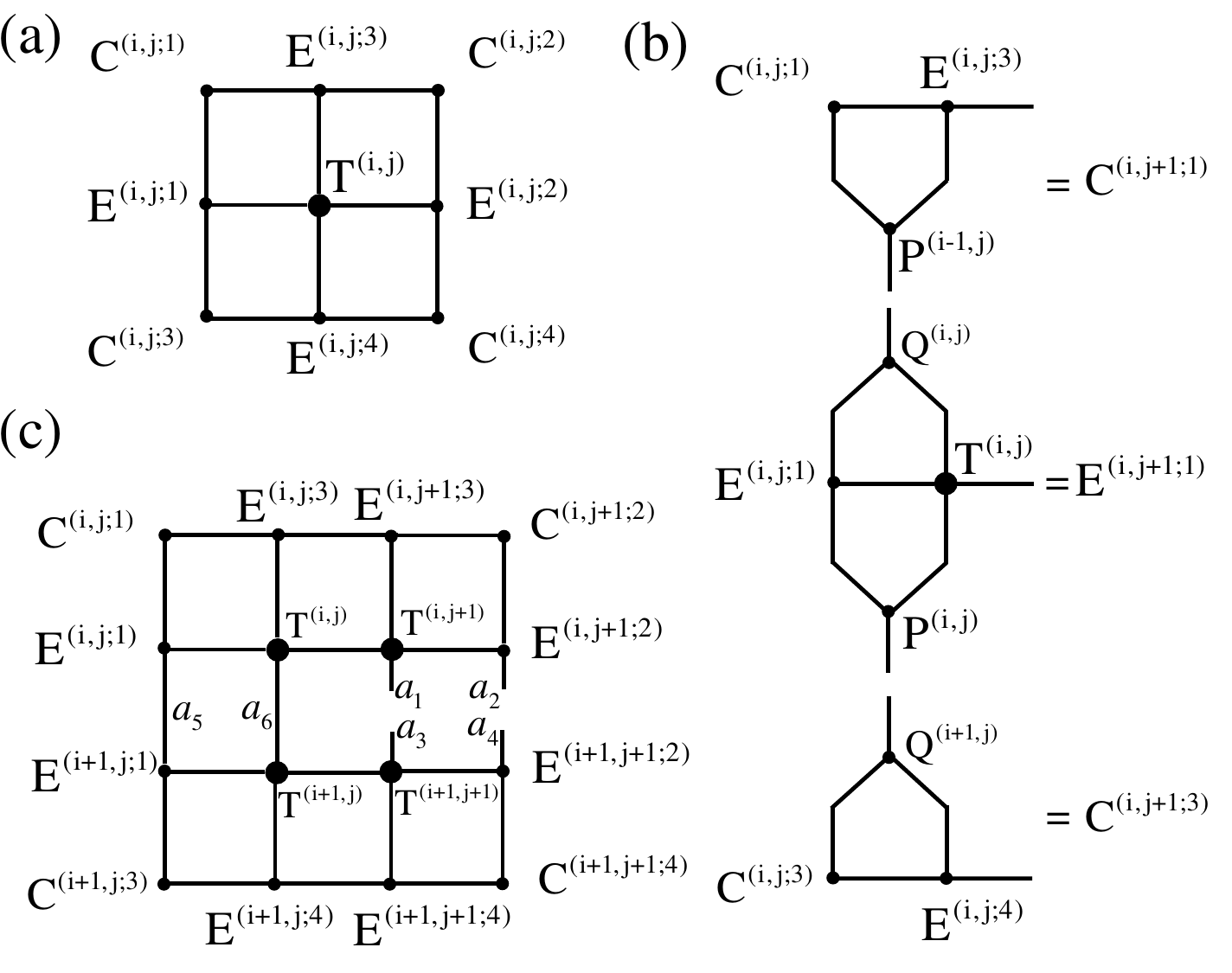}
	\caption{A sketch of the CTMRG algorithm used in this work, for a tensor network with a $4\times 4$ unit cell. The superscripts $i$ and $j$ should be taken modulo 4. (a) The environment of the local tensor located at the $i$ th row and $j$ th column in the unit cell. (b) The recursive relation between the environment of $T^{(i,j)}$ and $T^{(i+1,j)}$, in a left-move. (c) The $4\times 4$ cluster used to determine the transformation matrices used in (b). Here $P^{(i,j)}$ and $Q^{(i+1,j)}$ will be determined, as expressed in Eqs.~(\ref{Eq:PQ}-\ref{Eq:QRSVD}), and be inserted in the bonds denoted by $a_5a_6$. The resulting framework constitutes an optimal approximation of the bond density matrix $\rho$, which has subscripts $(a_1a_2,a_3a_4)$.}
	\label{Fig:CTMRG}
\end{figure}

    It should be noted that, by converting the reduced double-layer structure in Fig.~\ref{Fig:NTN}(a) to the nested single-layer structure in Fig.~\ref{Fig:NTN}(c), we have made an acceptable sacrifice to reduce the bond dimension $D^2$ to $Dd$: the size of the unit cell is doubled from $2\times 2$ to $4\times 4$. Therefore, to contract the tensor network by CTMRG, e.g., one has to employ the multisite version of the algorithm adapted to the nested structure. The specific version of the CTMRG algorithm used in this work, for a $4\times 4$ unit cell, is illustrated in Fig.~\ref{Fig:CTMRG}, which is a simple variant of the algorithm proposed in Ref.~\cite{Corboz-MulCTMRG}. 

     Specifically, suppose the distinct tensors in a unit cell are denoted as $\{T^{(i,j)}, i,j=0,1,2,3\}$, which are all rank-4 tensors with bond dimension $D$ \cite{Explain}. In the multisite CTMRG method, the environment of the tensor $T^{(i,j)}$ located at the $i$ th row and $j$ th column in the $4\times 4$ unit cell is represented as $\{C^{(i,j;\alpha)}, E^{(i,j;\beta)}\}$, where $\alpha$ and $\beta$ run from 1 to 4 and correspond to the four corner and four edge tensors for each $T^{(i,j)}$, as illustrated in Fig.~\ref{Fig:CTMRG}(a). Therefore, in this case we have 64 distinct corners and 64 edges in total, stored independently for each $T^{(i,j)}$ but updated in a correlated manner. For example, in Fig.~\ref{Fig:CTMRG}(b) we explicitly show the recursive relations between the environments of $T^{(i,j)}$ and $T^{(i,j+1)}$, which are used in a left-move step in the CTMRG iterations. 
     
     The core step of CTMRG is to find the transformation matrices $P^{(i,j)}$ and $Q^{(i,j)}$, which are of dimension $D\chi$ by $\chi$. Here $\chi$ is the so-called environment dimension, which is the dimension of the bonds belonging only to the environment tensors. Following the bond density matrix approach developed in Refs.~\cite{HOTRG, Corboz-MulCTMRG}, we always use a $2\times 2$ system surrounded by its environment to determine $P$ and $Q$, as illustrated in Fig.~\ref{Fig:CTMRG}(c), where the $4\times 4$ block is intentionally split into two parts from right. Mathematically, $P^{(i,j)}$ and $Q^{(i+1,j)}$ are constructed in the following
\begin{eqnarray}
P^{(i,j)} = R_s^{t}V\tilde{\Lambda}^{-1/2}, \quad Q^{(i+1,j)}=R_n^{t}U^*\tilde{\Lambda}^{-1/2} 
\label{Eq:PQ}
\end{eqnarray}    
where we have introduced the QR decomposition and singular value decomposition of two matrices $N$ and $S$,
\begin{eqnarray}
N = Q_nR_n, \quad S = Q_sR_s, \quad R_nR_s^{t} = U\Lambda V^{\dagger}
\label{Eq:QRSVD}
\end{eqnarray}
Here the superscript $t$ means matrix transpose. The $\tilde{\Lambda}$ in Eq.~(\ref{Eq:PQ}) is the truncated version (with dimension $\chi$) of the full singular value matrix $\Lambda$ in Eq.~(\ref{Eq:QRSVD}). $N$ is the block tensor with subscripts $(a_1a_2,a_5a_6)$ by contracting the eight tensors in the upper part of Fig.~\ref{Fig:CTMRG}(c), and $S$ is tensor with subscripts $(a_3a_4,a_5a_6)$ by contracting the lower part. It can be proved that by inserting the transformation matrices $P^{(i,j)}\left(Q^{(i+1,j)}\right)^{t}$ in the bonds labeled by $a_5a_6$ between $N$ and $S$ in Fig.~\ref{Fig:CTMRG}(c), the resulting network constitutes a truncated singular value decomposition of the bond density matrix $\rho_{a_1a_2,a_3a_4}$ which can be expressed as $\rho = NS^{t}$. 

In a single step of left move, for a $4\times 4$ unit cell, we need to determine four pairs of transformation matrices by performing Eqs.~(\ref{Eq:PQ}-\ref{Eq:QRSVD}) for each $i$ independently. And after that, one can use these transformations to update the related three environment tensors for each $i$ by performing the contractions illustrated in Fig.~\ref{Fig:CTMRG}(b). This finishes a single step of the left move. We need to perform such a step for each $j$, which means we have to perform four such steps in a single left move. Right, upward, and downward moves can be performed one by one similarly, and this completes a single CTMRG iteration. This iteration can be repeated, until all the spectra $\tilde\Lambda$s converge to some accepted accuracy.  
     
From Fig.~\ref{Fig:CTMRG} and the analysis described above, it is easily shown 
that if the environment dimension $\chi$ is set to the order of $D^2$ 
empirically \cite{NTN2017}, and no symmetry or other advanced techniques like 
partial matrix decompositions are employed, then the above CTMRG algorithm 
utilized to contract a nested tensor network with bond dimension $D$ has a 
memory cost about $D^6$ and a computational cost about $D^9$ ~\cite{YJKao-CTMRG, 
DNSheng-CTMRG}. This is more economical than the conventional reduced tensor 
network method as mentioned in Sec.~\ref{Sec:NTN} and discussed in detail in 
Ref.~\cite{NTN2017}. And this is the reason why we can extend our variational 
calculation equipped with automatic differentiation to $D = 9$ in this work.
As also discussed in Ref.~\cite{NTN2017}, as long as $\chi$ is sufficiently large, the CTMRG contraction of a nested tensor network (i.e., single-layer) should produce the same result as that of the corresponding reduced tensor network (i.e., double-layer). We verify this by studying the VBS phase of the Shastry-Sutherland model, and the result is shown in the Appendix.

\section{Results} 
\label{Sec:Results}
In this section, we apply the single-layer variational tensor network framework, 
as explained in the last section, to two prototypical antiferromagnetic quantum 
spin-1/2 models. Even though neither 
spin symmetries, such as U(1) and 
SU(2), nor GPU hardware acceleration are utilized in this work, the ground-state 
energy and the order parameters characterizing the quantum phases are obtained 
accurately by pushing $D = 9$. This demonstrates the power of the proposed 
single-layer framework.

\subsection{The Heisenberg model on a square lattice} 
\begin{figure}[!h]
	\centering
	\includegraphics[scale=0.35]{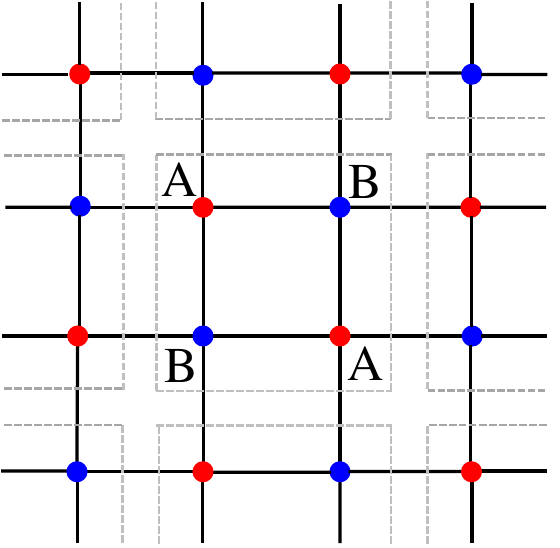}
	\caption{The PEPS ansatz used in this work to study the square-lattice Heisenberg model. Dashed lines separate the $2\times 2$ unit cells.}
	\label{Fig:Ans}
\end{figure}

The first model we studied is the antiferromagnetic spin-1/2 Heisenberg model on 
a square lattice, which was known to host a N\'{e}el antiferromagnetic 
ground state. To obtain such a state with a double sublattice structure, we use 
a compatible PEPS ansatz with $2\times 2$ unit cell containing only two distinct 
local tensors, as shown in Fig.~\ref{Fig:Ans}. 

\begin{figure}[!h]
	\centering
	\includegraphics[scale=0.16]{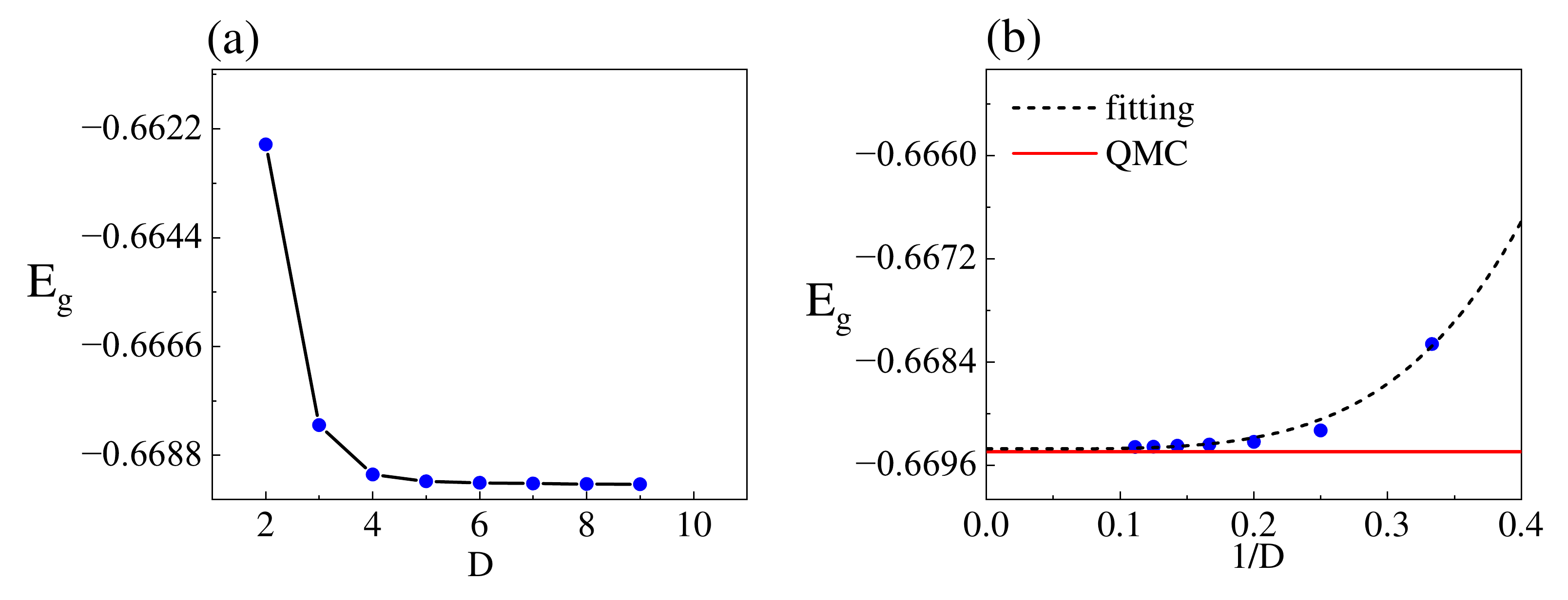}
	\caption{The obtained ground-state energy $E_g$ of the square lattice 
	Heisenberg model. (a) $E_g$ as a function of bond dimension $D$, with $D$ 
	ranging from 2 to 9. (b) A power-law fitting of data $E_g(1/D)$. The 
	extrapolation to the large-$D$ limit gives $E_g = -0.66941$. A quantum Monte 
	Carlo estimate from Ref.~\cite{HM-MC}, $E_g=-0.6694421$, is also shown as 
	red solid.}
	\label{Fig:HMEnergy}
\end{figure}

The estimated ground-state energy is plotted in Fig.~\ref{Fig:HMEnergy} as a 
function of $D$. It shows that the obtained energy systematically decreases as 
$D$ increases, verifying the variational nature of the framework. When $D = 9$, 
we obtain an estimate of $E_g=-0.6693896$, which is already very close to 
the 
value $E_g=-0.6694421$ in quantum Monte Carlo simulation \cite{HM-MC}. If we 
make a further power-law fitting of $E_g(1/D)$, then in the large-$D$ limit we 
obtain an extrapolated value of $E_g = -0.66941$, which deviates from the quantum 
Monte Carlo estimate only by the order of $10^{-5}$. 
	
\begin{figure}[!h]
	\centering
	\includegraphics[scale=0.18]{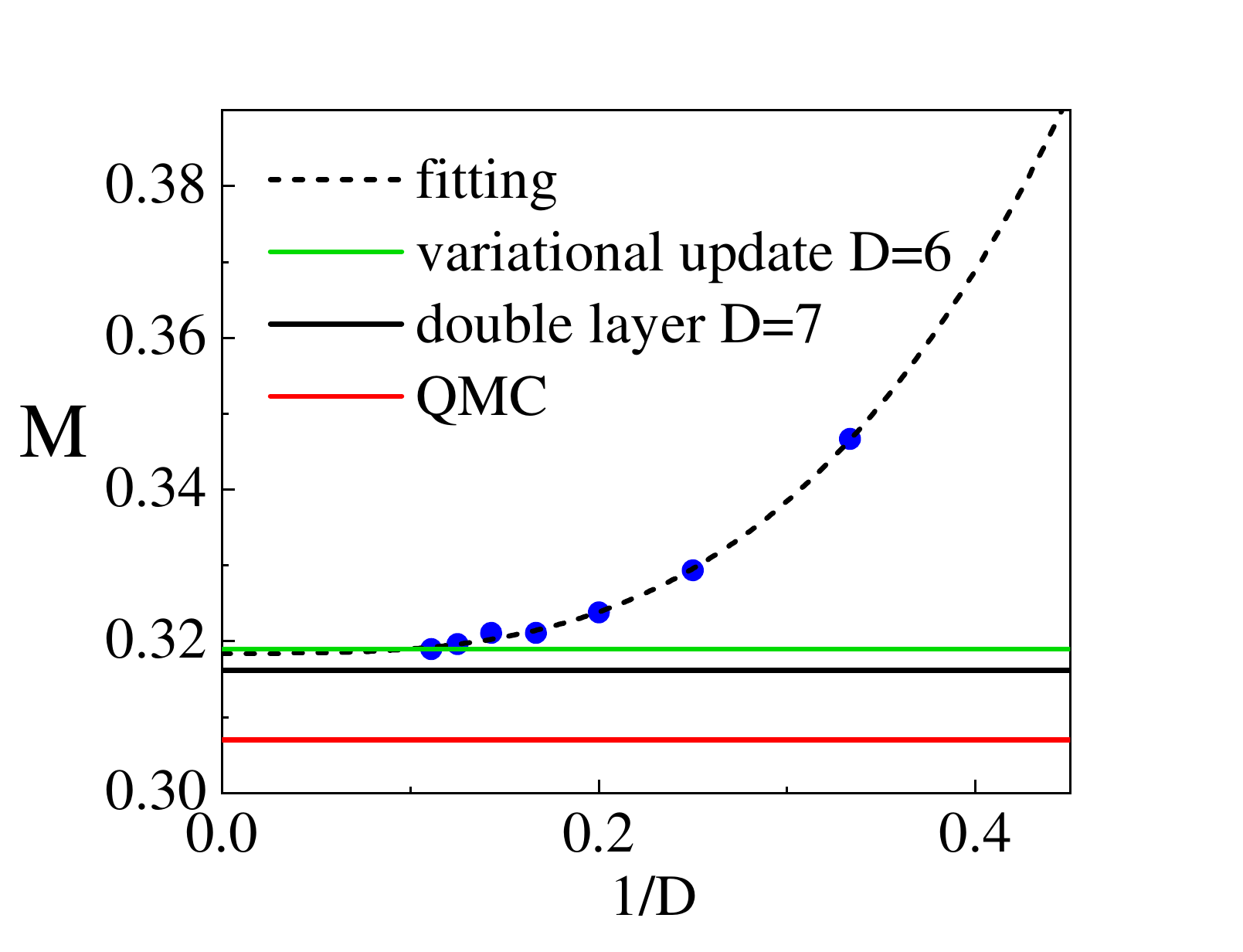}
	\caption{The obtained magnetization $M$ of the square lattice Heisenberg 
	model. A power-law fitting of data (dashed line) gives $M = 0.3159$ in the 
	large-$D$ limit. The solid red line marks the value $M = 0.307$ from a 
	quantum Monte Carlo estimate~\cite{HM-MC}. The solid black and green lines 
	mark the values of $M$ from a $D = 7$ double layer \cite{LW-PTNS2019} and a 
	$D=6$ variational update \cite{Corboz-vf} tensor network calculations.}
	\label{Fig:HMMag}
\end{figure}

As mentioned before, the ground state of this model is antiferromagnetically 
long-range ordered; therefore, we can use the magnetization $M$ defined below as 
an order parameter to characterize the ground state:
\begin{eqnarray}
M = \frac{1}{N}\sum_{i}|m_i|, \quad m_i = \left(\langle S^{(i)}_x\rangle, \langle S^{(i)}_y\rangle, \langle S^{(i)}_z\rangle\right)
\label{Eq:Mag1}
\end{eqnarray}	
where $S^{(i)}_\alpha$ is the $\alpha$ component of the spin operator defined on 
the $i$ th inequivalent site in the unit cell, $|m_i|$ is the length of the 
magnetic vector $m_i$, and $N$ is the total number of inequivalent sites in a 
unit cell. The obtained magnetization $M$ as a function of $D$ is plotted in 
Fig.~\ref{Fig:HMMag}. Similarly to $E_g$, the magnetization is also 
systematically lowered as $D$ increases.  When $D = 9$, $M$ is lowered to 
0.31897, and the extrapolated value in the large-$D$ limit obtained by a power 
fitting of the data $M(1/D)$ gives an estimate $M = 0.3159$, which is 
only slightly higher than the value $M = 0.307$ in quantum Monte Carlo 
simulation 
\cite{HM-MC}, but comparable to those from double-layer and variational update 
calculations~\cite{LW-PTNS2019,Corboz-vf}. What is more, we find that the local 
magnetization $m_i$ exhibits 
an explicit antiferromagnetic pattern indicated by staggered orientations. All 
these signatures are consistent with the N\'{e}el antiferromagnetic ground state.

\begin{figure}[!h]
	\centering
	\includegraphics[scale=0.16]{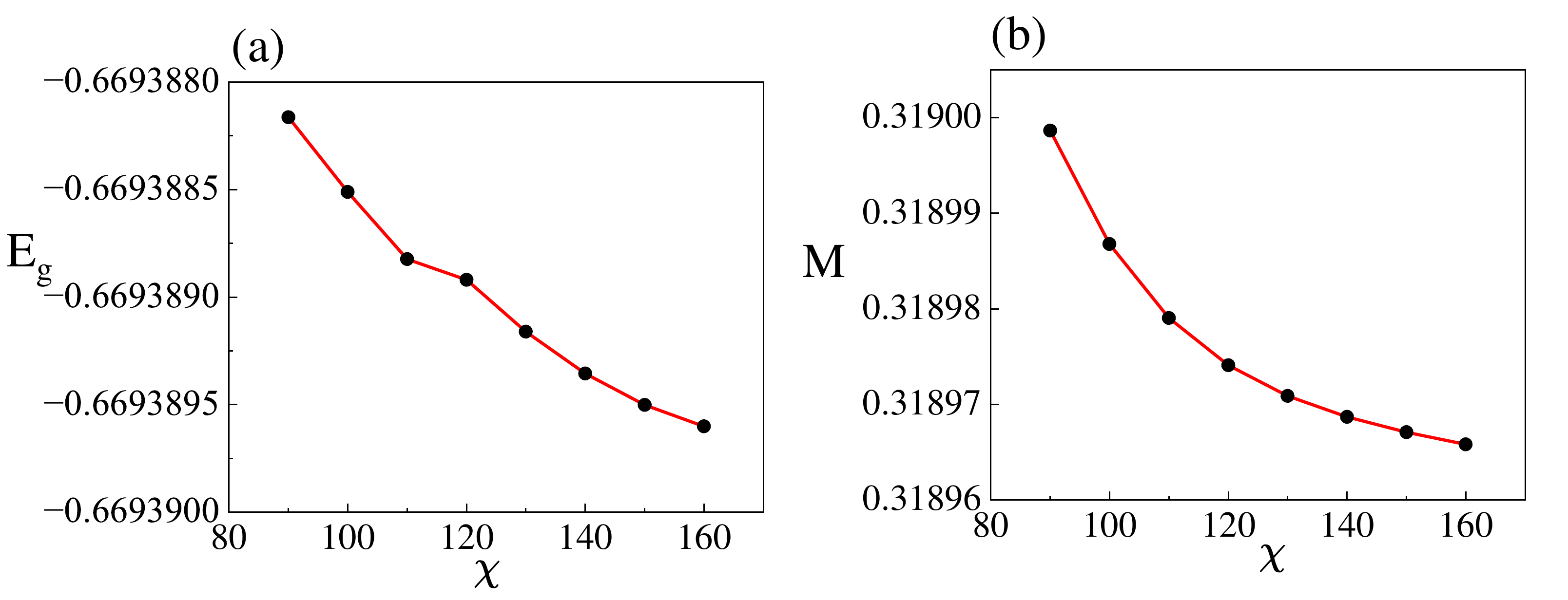}
	\caption{Convergence analysis of $E_g$ and $M$ with respect to the environment dimension $\chi$ used in CTMRG, for the square lattice Heisenberg model with $ D = 9$. (a) The ground-state energy $E_g$. (b) The magnetization $M$.}
	\label{Fig:HMChi}
\end{figure}

As described in Sec.~\ref{Sec:Method}, there are two important hyperparameters in variational tensor network calculations. One is the bond dimension $D$ of the PEPS ansatz as expressed in Eq.~(\ref{Eq:PEPS}), and the other is the environment dimension $\chi$ used in the CTMRG algorithms to evaluate $\langle\Psi|\Psi\rangle$ for a given $D$. Essentially, the calculation is variational with respect to $D$, but is not variational for $\chi$. Therefore, for a given $D$, the expectation values $E(D)$ and $M(D)$ are reliable only under the condition that they are converged with respect to $\chi$ as $\chi$ increases \cite{NTN2017}. Therefore, in this work, in order to ensure the convergence of $E$ and $M$ for each $D$, we always keep the environment dimension $\chi$ sufficiently large. A detailed analysis for $D = 9$ is shown in Fig.~\ref{Fig:HMChi}. It shows that by increasing $\chi$ to 160, both the energy and the magnetization are systematically improved, and the convergence error is only of the order of $10^{-6}$, which is already rather satisfactory for our purpose.

\subsection{The frustrated Shastry-Sutherland model}
\begin{figure}[!h]
	\centering
	\includegraphics[scale=0.25]{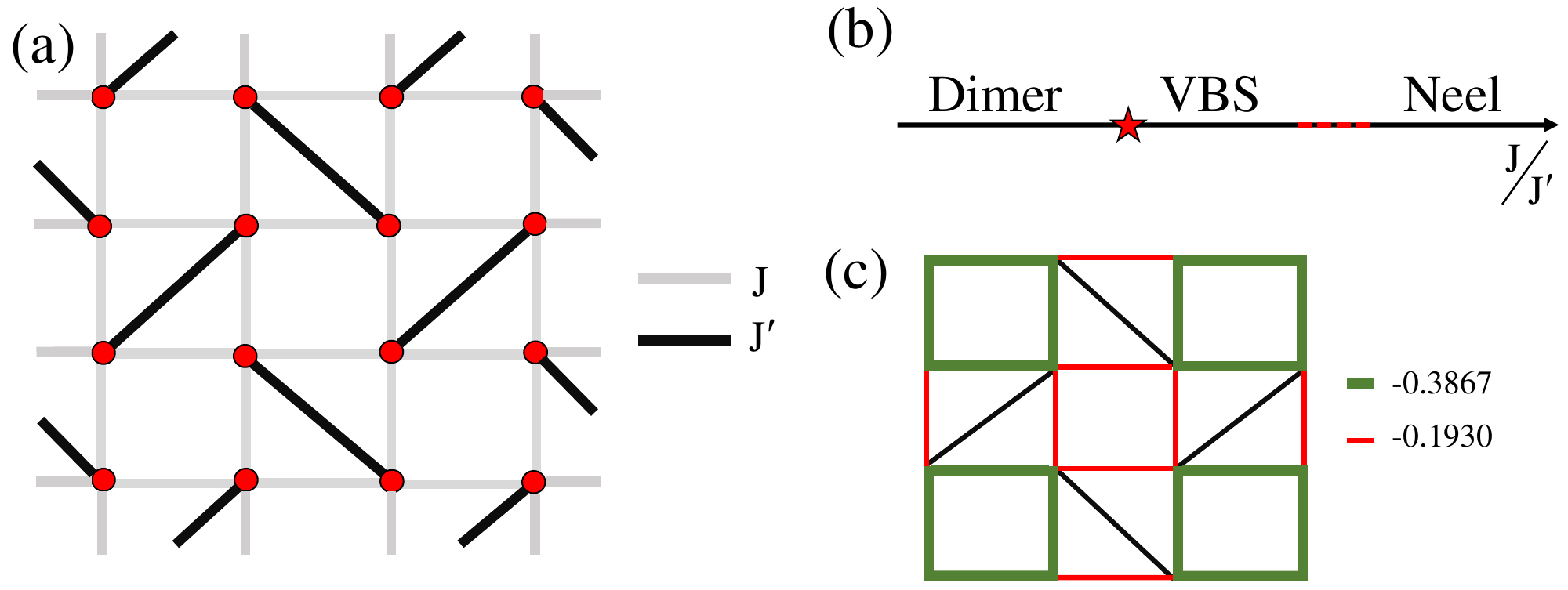}
	\caption{The Shastry-Sutherland model we studied in this work, as expressed 
	in Eq.~(\ref{Eq:SSM}). (a) The Shasty-Sutherland lattice. The grey bonds 
	constitute the underlying square lattice, and the coupling strength on these 
	bonds corresponds to $J$. The black bonds correspond to the coupling $J'$ 
	that forms a wheelchair pattern. (b) A sketch of the phase diagram 
	consisting of a dimer singlet, a plaquette valence bond solid (VBS), and a 
	N\'{e}el antiferromagnetic ground state with varying $J/J'$. The nature of 
	the plaquette VBS to the antiferromagnetic transition is under debate, with  
	the possiblity of an intervening quantum spin liquid as marked by the red 
	dashed region. (c) The bond correlation 
	$\langle\spins_i\cdot\spins_j\rangle$ in the obtained ground state at $J/J' 
	= 0.7$ in this work, consistent with the 
	empty-plaquette VBS state \cite{NingXi-SSM2023}. }
	\label{Fig:SSM}
\end{figure}

The second model we focused on in this work is the spin-1/2 Shastry-Sutherland model defined by the following Hamiltonian
\begin{equation}
H = J\sum_{\langle i,j \rangle} \spins_i\cdot\spins_j + J^{\prime}\sum_{\langle\langle i,j \rangle\rangle} \spins_i\cdot\spins_j, \quad J>0,~J'>0
\label{Eq:SSM}
\end{equation}
where $\spins_i$ is the three-component spin operator defined on $i$ th lattice 
point. A sketch of the Shastry-Sutherland lattice is illustrated in 
Fig.~\ref{Fig:SSM}(a). From the view of the underlying square lattice, one can 
see that in Eq.~(\ref{Eq:SSM}), $J$ denotes the nearest-neighboring 
antiferromagnetic coupling, and $J'$ denotes the next-nearest-neighbor 
coupling defined in only one set of squares and arranged as a wheelchair 
pattern. This model draws attention in recent years \cite{Corboz-SSM, 
Corboz-SSMH, NingXi-SSM2023, WYL-PRL2024} because of the stabilization of a 
quantum disordered plaquette valence bond solid ground state in the intermediate 
$J/J^\prime$ regime of the phase diagram, as well as possible realization of 
a deconfined quantum critical point \cite{DQCP-Senthil} and/or quantum spin 
liquid~\cite{Corboz-QSLSSM2025, WYL-PRL2024, LingWang-SSM} 
relevant to some interesting experimental results in real materials 
\cite{SCBO-Science}.

An illustration of the phase diagram is also shown in Fig.~\ref{Fig:SSM}(b). 
When $J$ dominates, the ground state of this model is a N\'{e}el 
antiferromagnetic state, while when $J'$ dominates, the two spins coupled by 
$J'$, as indicated by the black bonds in Fig.~\ref{Fig:SSM}(a), form a dimer 
singlet, and the ground state of the system is a direct product of these dimer 
singlets. In the intermediate regime, the ground state is a plaquette valence 
bond solid. However,
there is much controversy over 
the competition between the full-plaquette and empty-plaquette VBS patterns 
\cite{Corboz-SSM, NingXi-SSM2023, SCBO-Science}, the nature of the phase 
transition between the N\'{e}el and VBS phases \cite{NingXi-SSM2023, 
SCBO-Science}, and the possible existence of an intervening quantum spin liquid 
regime \cite{Corboz-QSLSSM2025, WYL-PRL2024, LingWang-SSM}. In this work, using 
the proposed single-layer variational framework and setting $J' = 1$, we focus 
on two parameter settings: the possible VBS phase with $J = 0.7$, and the 
antiferromagnetic phase with $J = 0.8$. 

\begin{figure}[!h]
	\centering
	\includegraphics[scale=0.16]{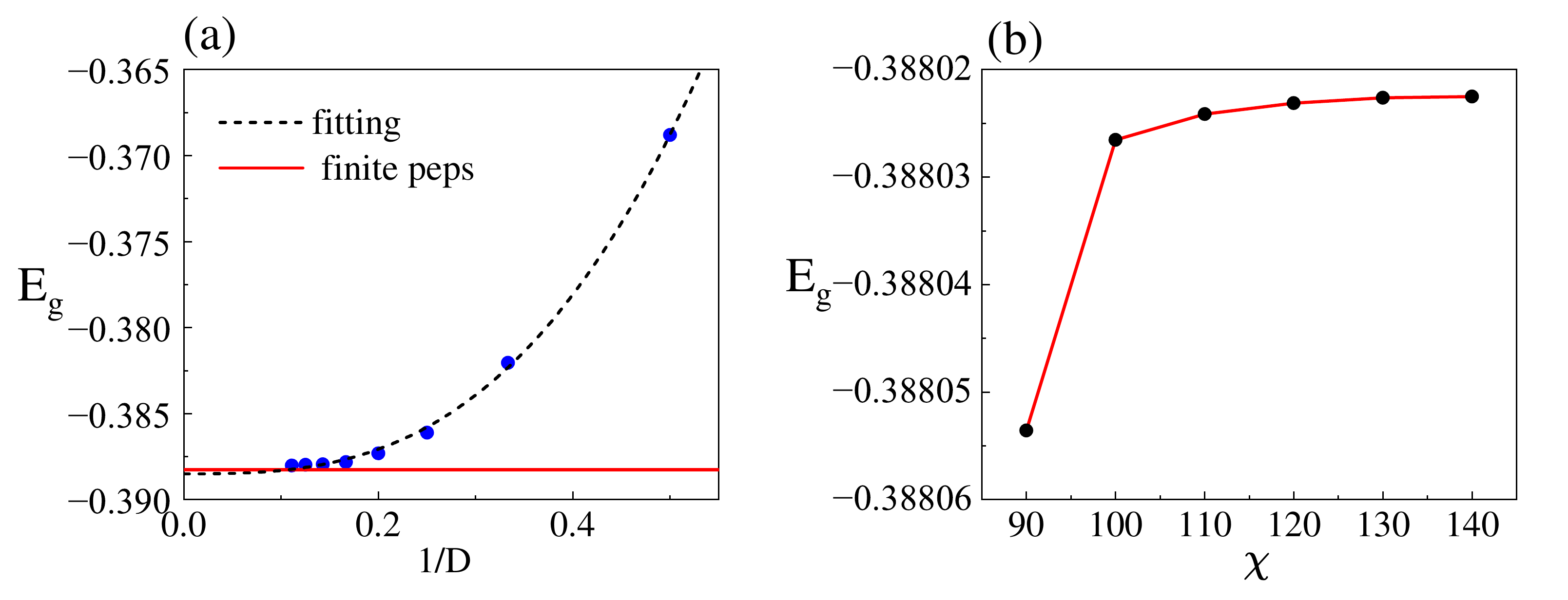}
	\caption{The obtained ground-state energy $E_g$ of the Shastry-Sutherland model at $J = 0.7$. (a) $E_g$ as a function of bond dimension $D$, with $D$ ranging from 2 to 9. A power-law fitting of data $E_g(1/D)$ is indicated by a black dashed line. For comparison, an extrapolation from finite PEPS calculations \cite{WYL-PRL2024} is also plotted as a red solid line. (b) The convergence analysis of $E_g(\chi)$ for $D = 9$. }
	\label{Fig:SSM-E1}
\end{figure}

In Fig.~\ref{Fig:SSM-E1}, we focus on $J = 0.7$, and plot the obtained 
ground-state energy $E_g$ as a function of $D$. The variational feature remains 
pronounced, i.e., the energy systematically improves as $D$ becomes larger. When 
$D = 9$, we have $E_g = -0.3880225$. If we fit the data $E_g(1/D)$ and 
extrapolate to the large-$D$ limit, we obtain an estimate $E_g = -0.3885$ 
[black dashed line in Fig.~\ref{Fig:SSM-E1}(a)], which is consistent with the previous 
PEPS from finite systems \cite{WYL-PRL2024}. To make the result convincing, we 
analyze the energy for $D = 9$ with $\chi$ ranging from 90 to 140 in 
Fig.~\ref{Fig:SSM-E1}(b). It is clear that the energy has almost fully converged 
with a precision of about $10^{-6}$. 

\begin{figure}[!h]
	\centering
	\includegraphics[scale=0.16]{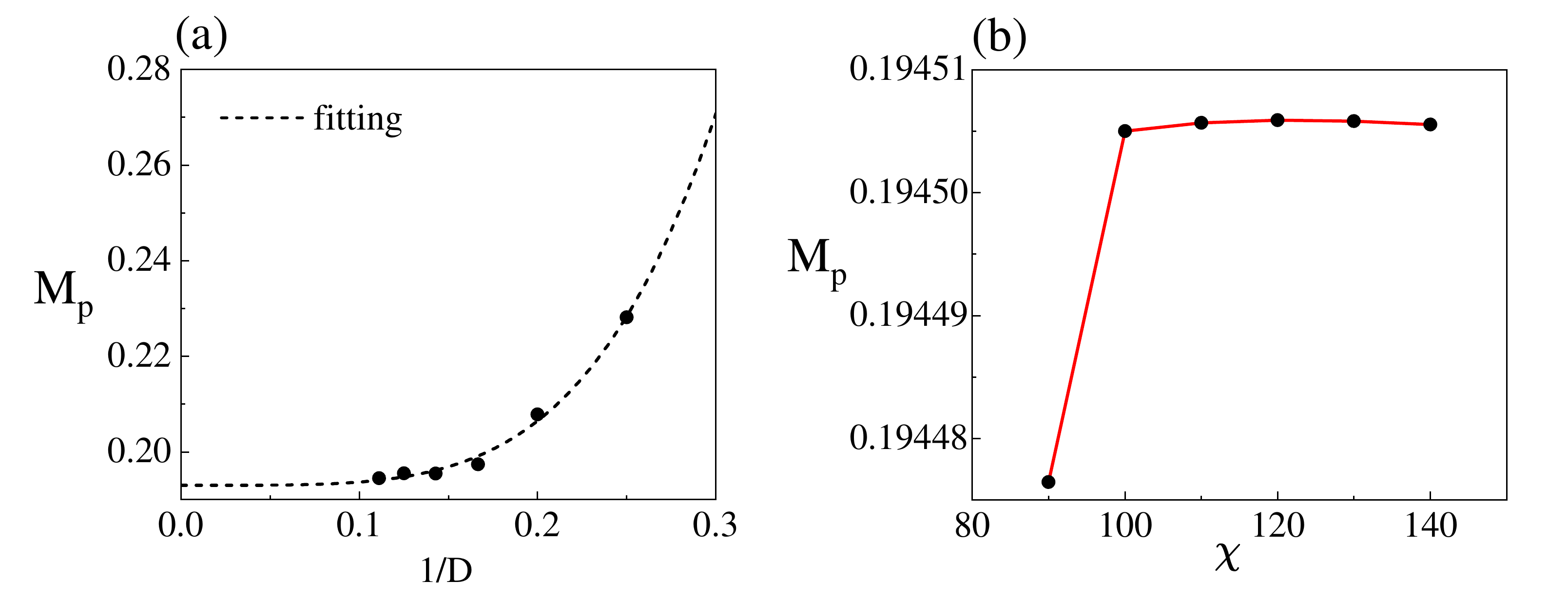}
	\caption{The obtained order parameter $M_p$, defined in Eq.~(\ref{Eq:Mag2}), of the Shastry-Sutherland model at $J = 0.7$. (a) $M_p$ as a function of bond dimension $D$, with $D$ ranging from 4 to 9. A power-law fitting of data $M_p(1/D)$ is also shown and denoted by a black dashed line. (b) The convergence analysis of $M_p(\chi)$ for $D = 9$.}
	\label{Fig:SSM-OP1}
\end{figure}

Further analysis shows that the obtained ground state is an empty-plaquette VBS state, which is illustrated in Fig.~\ref{Fig:SSM}(c). To show this, first we introduce an order parameter $M_p$ defined as
\begin{eqnarray}
M_p = \frac{1}{4}\Big|\sum_{\langle i,j\rangle\in\square_A}\langle\mathbf{S}_i\cdot\mathbf{S}_j\rangle - \sum_{\langle i,j\rangle\in\square_B}\langle\mathbf{S}_i\cdot\mathbf{S}_j\rangle\Big|
\label{Eq:Mag2}
\end{eqnarray}
where the two $\sum$ sum over all the nearest-neighbor bond energies in  inequivalent empty squares 
		$\square_{A}$ and $\square_{B}$ [olive and red plaquettes illustrated in 
		Fig.~8(c)].
In Fig.~\ref{Fig:SSM-OP1}(a), we plot the obtained $M_p$ as a function of $D$. It 
shows clearly that the order parameter converges gradually as $D$ increases, and 
when $D = 9$ we obtain $M_p = $ 0.1945. By extrapolating to the large-$D$ limit, we 
obtain a final estimate $M_p = $ 0.1930, which is a strong indication that the 
ground state breaks the $Z_2$ symmetry between two inequivalent squares. The 
convergence with respect to $\chi$ plotted in Fig.~\ref{Fig:SSM-OP1}(b) 
indicates that the symmetry breaking is robust and is not a finite-$\chi$ 
effect. To show the detailed pattern and distinguish the two possible VBS states 
\cite{NingXi-SSM2023, Corboz-SSM}, we plot the bond correlation 
$\langle\mathbf{S}_i\cdot\mathbf{S}_j\rangle$ for each bond in 
Fig.~\ref{Fig:SSM}(c). It shows that the correlation in squares without $J'$ 
couplings is much stronger than that in squares with $J'$ couplings, consistent 
with the empty-plaquette VBS state.  Furthermore, we find the obtained ground 
state has a tiny magnetization of $\sim 10^{-4}$, suggesting it probably does 
not break the SU(2) spin symmetry. Combining these features, we can convincingly 
conclude that the ground state is an empty-plaquette VBS state at $J = 0.7$, 
consistent with previous studies \cite{NingXi-SSM2023}.
 
\begin{figure}[!h]
	\centering
	\includegraphics[scale=0.16]{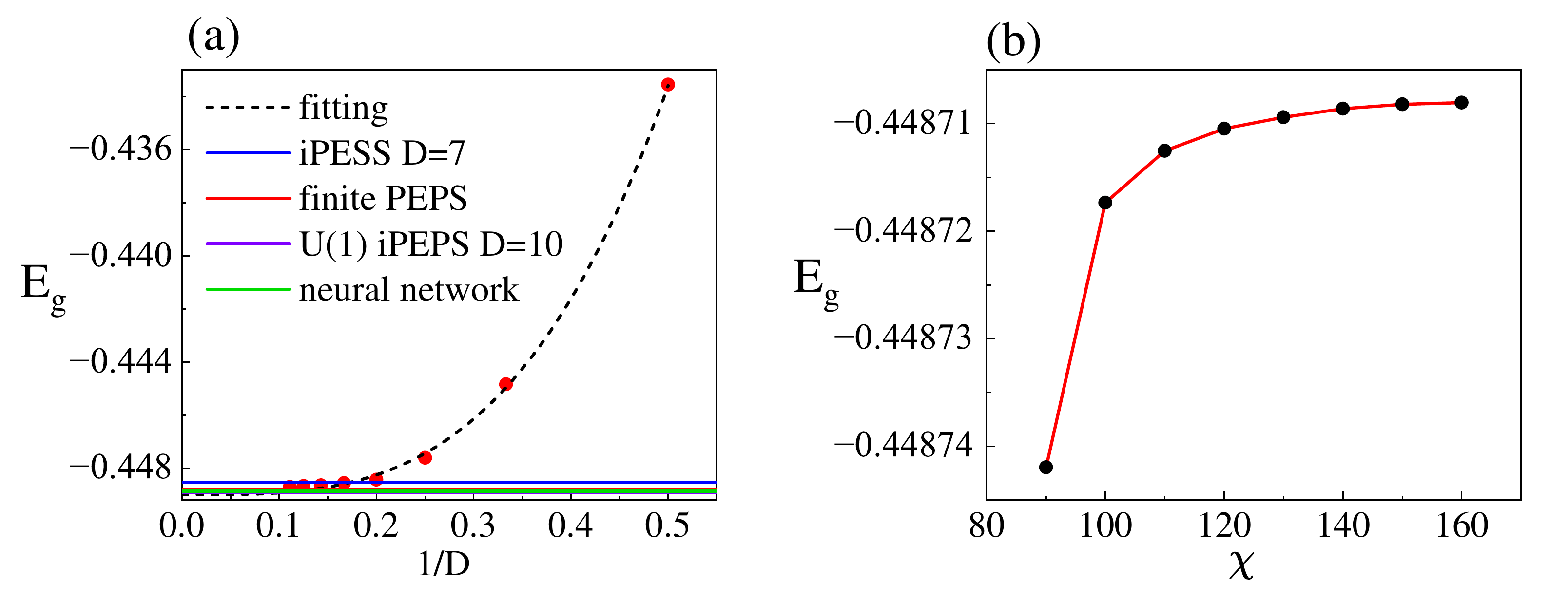}
	\caption{The obtained ground-state energy $E_g$ of the Shastry-Sutherland model at $J = 0.8$. (a) $E_g$ as a function of bond dimension $D$, with $D$ ranging from 2 to 9. A power-law fitting of data $E_g(1/D)$ is indicated by a black dashed line. For comparison, an iPESS study \cite{NingXi-SSM2023}, an infinite PEPS study employing U(1) symmetry \cite{Corboz-QSLSSM2025}, a neural network study \cite{NQS} and an extrapolation from finite PEPS calculations \cite{WYL-PRL2024}, are also included as solid lines. (b) The convergence analysis of $E_g(\chi)$ for $D = 9$.}
	\label{Fig:SSM-E2}
\end{figure}

For $J = 0.8$, we plot the ground-state energy in Fig.~\ref{Fig:SSM-E2}. The behavior is similar to that for $J = 0.7$, and the energy systematically decreases and gradually converges as $D$ becomes larger. When $D = 9$, we have $E_g = -0.448708$. This is already very close to the previous tensor network studies, such as an infinite projected entangled simplex state (iPESS) study \cite{NingXi-SSM2023}, an infinite PEPS study employing U(1) spin symmetry \cite{Corboz-QSLSSM2025}, a neural network study \cite{NQS} and an extrapolation from finite PEPS calculation \cite{WYL-PRL2024}. These estimate are also included in Fig.~\ref{Fig:SSM-E2} for comparison. Extrapolation to the large-$D$ limit via a simple power-law fitting yields an estimate of $E_g = -0.4490$, which is more reliable. With $\chi$ up to 160, Fig.~\ref{Fig:SSM-E2} shows that the convergence precision of the ground-state energy for $D=9$ has reached the order $10^{-6}$, indicating that the obtained result is convincing.

\begin{figure}[!h]
	\centering
	\includegraphics[scale=0.16]{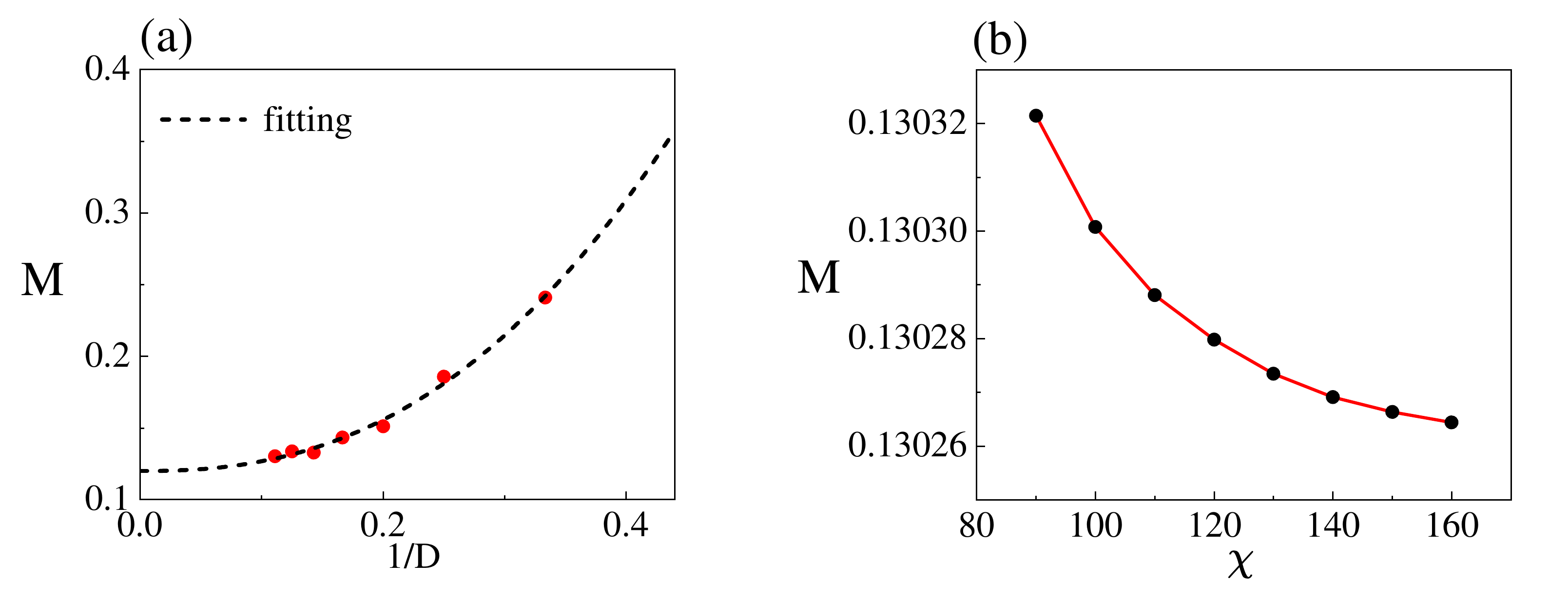}
	\caption{The obtained order parameter $M$, defined in Eq.~(\ref{Eq:Mag1}), of the Shastry-Sutherland model at $J = 0.8$. (a) $M$ as a function of bond dimension $D$, with $D$ ranging from 3 to 9. A power-law fitting to the data $M(1/D)$ is also shown in black dashed. (b) The convergence analysis of $M(\chi)$ for $D = 9$.}
	\label{Fig:SSM-OP2}
\end{figure}

In Fig.~\ref{Fig:SSM-OP2}, we plot the magnetization $M$ of the obtained ground 
state as a function of $D$. It shows that though $M$ becomes weaker and weaker 
as $D$ increases, it remains finite and nonzero. For example, when $D = 9$, we 
still obtain a significant magnetization of about 0.1303, and an extrapolation 
in the large-$D$ limit still gives a nonzero estimate of about $M = 0.12$.  
This is neither an artificial finite-$\chi$ effect, since by pushing $\chi$ to 
160, the convergence precision of $M$ for $D = 9$ has also reached the order 
$10^{-6}$, as clearly shown in Fig.~\ref{Fig:SSM-OP2}(b). This is unambiguously 
consistent with the fact that $J = 0.8$ falls in the antiferromagnetic phase.

\section{Discussions}
\label{Sec:Outlook}
In this work, we propose a single-layer framework that integrates the nested tensor network method into the variational tensor network calculations leveraging automatic differentiation. This framework can reduce the memory and computational costs by two and three orders of $D$ compared to traditional variational calculations, thereby enabling $D$ to be larger and achieving better accuracy. The validity of the framework is demonstrated through two prototypical models: the square-lattice Heisenberg model and the Shastry-Sutherland model. Even without GPU hardware acceleration or symmetry considerations, we can extend $D$ to 9, yielding accurate ground-state energies and consistent order parameters with previous studies. In particular, for the Shastry-Sutherland model, we confirm an intermediate empty-plaquette VBS phase in the ground-state phase diagram. The convergence with respect to $D$ and $\chi$ is systematically analyzed, and nearly complete convergence is observed even for the largest $D$. We believe this framework could enable large-scale variational tensor network calculations in more complex quantum systems, such as superconductivity \cite{TNStJ, TNSHubbard}, excited states, and dynamical studies \cite{ADPEPS, LY-PRB2025} in the future. 

In our calculations, the single-layer tensor network with nested structure is contracted by the CTMRG algorithm. It is known that the idea of nested tensor network can be performed not only in CTMRG \cite{YJKao-CTMRG, DNSheng-CTMRG}, but also in other tensor contraction algorithms, such as time-evolving block decimation \cite{NTN2017, Vidal-TEBD, Orus-TEBD}, and coarse-graining renormalization group. Therefore, correspondingly, the single-layer framework in this work can also be realized in other algorithms than CTMRG. The extra acceptable sacrifice is probably the same, i.e., the size of the unit cell of the single-layer nested tensor network may be doubled, as shown in Fig.~\ref{Fig:NTN}(c) in this work.

The bond dimension $D$ can be further increased by incorporating additional advanced techniques into this framework. For example, in this work we did not consider the spin symmetry, but one can leverage the U(1) symmetry of the two models to improve efficiency. For the square lattice Heisenberg model in this work, one can utilize a local spin rotation $-i\sigma_y$ on one set of lattice locations \cite{Hasik-Sci2021, LY-PRB2025} to reduce the $4\times 4$ unit cell to a $2\times 2$ unit cell in the generated nested tensor network, which can simplify the calculations shown in Fig.~\ref{Fig:NTN}(c). Even for the Shastry-Sutherland model, this strategy can accelerate the convergence in variational calculations. Moreover, one can use a checkpoint \cite{LW-PTNS2019} or a fixed-point technique to reduce the length of the chain, i.e., Eq.~(\ref{Eq:AD}) used in the back propagation, by calculating the gradient from a given or an approximate fixed-point environment \cite{LW-fixedMPS}. The latter strategy is similar to the finite-system approximation \cite{SRG2010} or to choosing a better initial value for the CTMRG environment and can effectively reduce memory and computational costs in automatic differentiation calculations. Finally, at a purely computational level, one can use partial singular value decomposition, randomized decomposition \cite{RSVD} in Eq.~(\ref{Eq:QRSVD}), or employ a more economical QR decomposition \cite{Corboz-QR2025} instead, to reduce computational cost further.

As reviewed in Sec.~\ref{Sec:NTN}, the nested tensor network method treats the bra $\langle\Psi|$ and ket $|\Psi\rangle$ independently when contracting $\langle\Psi|\Psi\rangle$, thus the Hermiticity of the environment (for the reduced tensor) \cite{Corboz-Z2} can be easily broken. Fortunately, the Hermiticity can be gradually restored by increasing the environment dimension $\chi$, and could be further improved by considering the hidden $Z_2$ symmetry \cite{Corboz-Z2}, more delicately designed iterations \cite{Schmoll-PRB2025}, and smarter variation with better convergence \cite{ZhangXY-PRB2026}, etc. The restoration of the Hermiticity is expected to improve the accuracy of the results.

It is worth noting that several recent studies also involve single-layer tensor network states in the literature. For example, in Ref.~\cite{Schmoll-PRB2025}, Jan \emph{et al.} proposed a similar framework called the split-CTMRG algorithm, which is also based on the nested tensor network structure revealed in Ref.~\cite{NTN2017}. However, the contraction scheme is particularly designed and differs from what we proposed in Sec.~\ref{Sec:SLF}. In Ref.~\cite{ZGM-PRB2025}, Xu \emph{et al.} extended the idea of the nested structure to a triple-layer tensor network and applied it to the three-dimensional Ising model successfully, while earlier in Ref.~\cite{YLP-PRB2023}, Yang \emph{et al.} proposed a new nesting structure that is more natural and symmetric than the one proposed in Ref.~\cite{NTN2017} and was applied successfully to the same model. Probably, these single-layer proposals can be further improved by combining their distinct advantages, which we would like to leave as a future direction. 

\section*{Acknowledgments.} 
We are grateful to Professor Tao Xiang and Ning Xi for helpful discussions, and to Jian-Gang Kong for reminding us of Ref.~\cite{Schmoll-PRB2025}. This work was supported by the National R\&D Program of China (Grants No. 2023YFA1406500 and No. 2024YFA1408604), the National Natural Science Foundation of China (Grants No. 12274458,and No. 12334008). 

\section*{Appendix}
\begin{figure}[!h]
	\centering
	\includegraphics[scale=0.16]{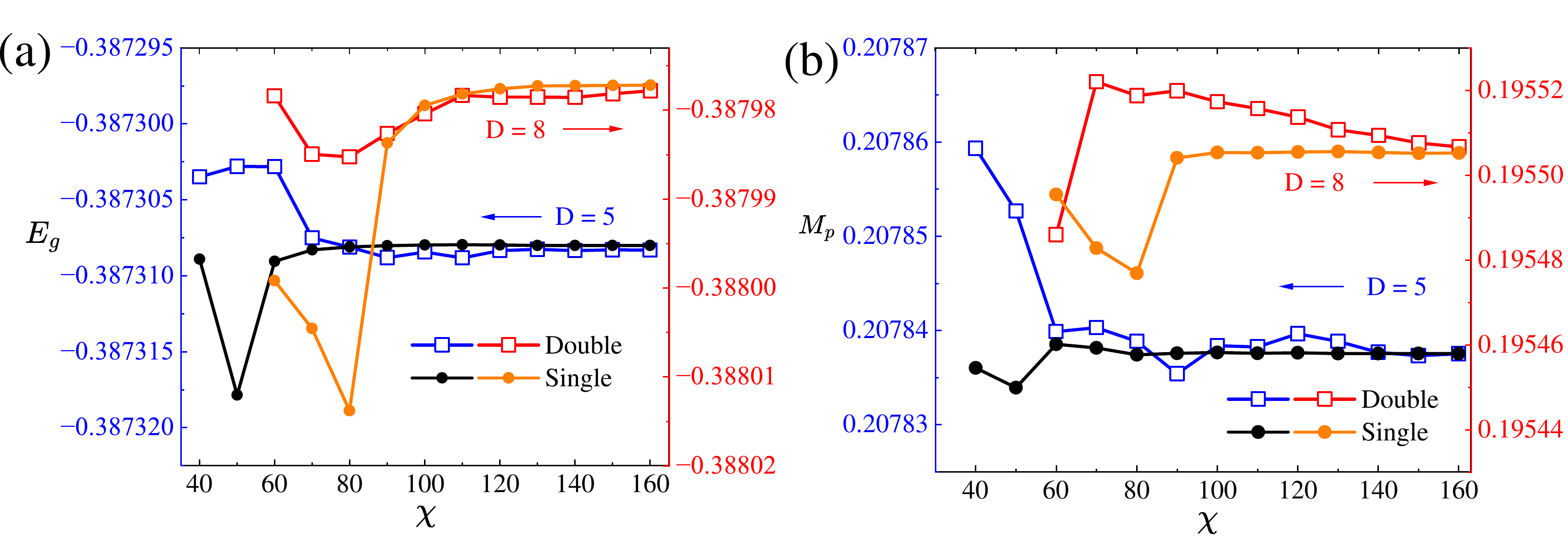}
	\caption{Comparison of (a) the ground-state energy $E_g$ and (b) the order parameter $M_p$ of the Shastry-Sutherland model at $J = 0.7$, between the two methods, i.e., single-layer CTMRG (nested, marked by dots) and double-layer CTMRG (reduced, marked by squares)}.
	\label{Fig:SM-Compare}
\end{figure}

In this appendix, using the Shastry-Sutherland model as an example, we compare results from the single-layer CTMRG and the corresponding double-layer CTMRG. We choose $J = 0.7$ (i.e., the plaquette VBS phase), and the wave functions are obtained from the single-layer variational calculations proposed in the main text. The results are plotted in Fig.~\ref{Fig:SM-Compare}. It shows that for both $D = 5$ and $D = 8$, the single-layer contraction can produce the same results as double-layer contraction as long as $\chi$ is sufficiently large (e.g., $\chi=160$ in this case). Moreover, at least for this model, the single-layer contraction leads to better convergence. This conclusion is similar to the Kagome Heisenberg model tested in other contraction frameworks \cite{NTN2017}.

\begin{thebibliography}{99}
\bibitem{XiangBook} T. Xiang, \emph{Density Matrix and Tensor Network Renormalization} (Cambridge University Press,Cambridge 2023). 
\bibitem{SimonBook} S. Montangero, \emph{Introduction to Tensor Network Methods} Springer (Switzerland, 2018).
\bibitem{PEPS2004} F. Verstraete, and J. I. Cirac, Renormalization algorithms for quantum-many body systems in
two and higher dimensions, arXiv:cond-mat/0407066.
\bibitem{OrusNatRev} R. Orus, Tensor networks for complex quantum systems, Nat. Rev. Phys. \textbf{1}, 538 (2019)
\bibitem{RanBook} S.-J. Ran, E. Tirrito, C. Peng, X. Chen, L. Tagliacozzo, G. Su, and M. Lewenstein, \emph{Tensor Network Contractions}, Springer (Open, 2020).
\bibitem{Orus2014} R. Orus, Tensor networks for complex quantum systems, Annals of Phys. \textbf{349}, 117 (2014).
\bibitem{OptBook} J. Nocedal, S. J. Wright, \emph{Numerical Optimization}, Springer (New York, 1999).
\bibitem{KagomePRL2017}  H. J. Liao, Z. Y. Xie, J. Chen, Z. Y. Liu, H. D. Xie, R. Z. Huang, B. Normand, T. Xiang, Gapless spin-liquid ground state in the kagome antiferromagnet, Phys. Rev. X \textbf{118}, 137202 (2017).
\bibitem{J1J2-LWY} W.-Y. Liu, D. Poilblanc, S.-S. Gong, W.-Q. Chen, Z.-C. Gu, Tensor network study of the spin 1/2 square-lattice model: Incommensurate spiral order, mixed valence-bond solids, and
multicritical points, Phys. Rev. B \textbf{109}, 235116 (2024).
\bibitem{NingXi-SSM2023} N. Xi, H. Chen, Z. Y. Xie, R. Yu, Plaquette valence bond solid to antiferromagnet transition and deconfined quantum critical point of the Shastry-Sutherland model, Phys. Rev. B \textbf{107}, L220408 (2023)




\bibitem{LY-PRB2025} Y. Liu, S. Shao, S. He, Z. Y. Xie, J.-W. Mei, H.-G. Luo, and J. Zhao, Quantum dynamics in a spin-1/2 square lattice $J_1-J_2-\delta$ altermagnet, Phys. Rev. B \textbf{111}, 245117 (2025).
\bibitem{TNStJ} Z. T. Xu, Z. C. Gu, and S. Yang, Competing orders in the honeycomb lattice t-J model, Phys. Rev. B \textbf{108}, 035144 (2023).
\bibitem{TNSHubbard} P. Corboz, Improved energy extrapolation with infinite projected entangled-pair states applied to
the two-dimensional Hubbard model, Phys. Rev. B \textbf{93}, 045116 (2016).
\bibitem{Corboz-MulCTMRG} P. Corboz, T. M. Rice, and M. Troyer,  Competing states in the t-J model: Uniform d-wave
state versus stripe state, Phys. Rev. Lett. \textbf{113}, 046402 (2014).
\bibitem{TNStopo} N. Schuch, D. Poilblanc, J. I. Cirac, D. Perez-Garcia, Topological order in the
projected entangled-pair states formalism: Transfer operator and boundary Hamiltonians, Phys. Rev. Lett. \textbf{111}, 090501 (2013).
\bibitem{CSL} R. Wang, T. Yang, Z. Y. Xie, B. Wang, X. C. Xie, Susceptibility indicator for chiral
topological orders emergent from correlated fermions, Phys. Rev. B \textbf{109}, L241113 (2024).
\bibitem{CMPS} F. Verstraete, and J. I. Cirac, Continuous matrix product states for quantum fields, Phys. Rev. Lett. \textbf{104}, 190405 (2010).
\bibitem{CTNS} A. Tilloy, and J. I. Cirac, Continuous tensor network states for quantum fields, Phys. Rev. X \textbf{9}, 021040 (2019).
\bibitem{QCmap} M. Suzuki, Relationship between d-dimensional quantal spin systems and (d+1)-dimensional
Ising systems: Equivalence, critical exponents and systematic approximants of the partition
function and spin correlations, Prog. Theor. Phys. \textbf{56}, 1454 (1976).
\bibitem{HOTRG} Z. Y. Xie, J. Chen, M. P. Qin, J. W. Zhu, L. P. Yang, T. Xiang, Coarse-graining
renormalization by higher-order singular value decomposition, Phys. Rev. B \textbf{86}, 045139 (2012).




\bibitem{SU} H. C. Jiang, Z. Y. Weng, and T. Xiang, Accurate determination of tensor network state of
quantum lattice models in two dimensions, Phys. Rev. Lett. \textbf{101}, 090603 (2008)
\bibitem{CU} L. Wang, F. Verstraete, Cluster update for tensor network states, arXiv:1110.4362.
\bibitem{FU} M. Lubasch, J. I. Cirac, and M. C. Banuls, Algorithms for finite projected entangled pair states, Phys. Rev. B \textbf{90}, 064425 (2014).
\bibitem{FFU} H. N. Phien, J. A. Bengua, H. D. Tuan, P. Corboz, R. Orus, Infinite projected entangled pair
states algorithm improved: Fast full update and gauge fixing, Phys. Rev. B \textbf{92}, 035142 (2015).
\bibitem{ADreview} C. C. Margossian, A review of automatic differentiation and its efficient implementation, WIREs
Data Min. Knowl. Discovery \textbf{9}, e1305 (2019).
\bibitem{NeuralBook} I. Goodfellow, Y. Bengio, and A. Couville, \emph{Deep Learning}, (MIT Press, Cambrige, MA, 2016).
\bibitem{LW-PTNS2019} H.-J. Liao, J.-G. Liu, L. Wang, and T. Xiang, Differentiable programming tensor networks, Phys. Rev. X \textbf{9}, 031041 (2019).
\bibitem{Pytorch} The official website of PyTorch is at \href{https://pytorch.org}{https://pytorch.org}.
\bibitem{Zygote} The website of Zygote is at \href{https://fluxml.ai/Zygote.jl}{https://fluxml.ai/Zygote.jl}.

\bibitem{PEPSComplex} N. Schuch, M. M. Wolf, F. Verstraete, J. I. Cirac, Computational complexity of projected
entangled pair states, Phys. Rev. Lett. \textbf{98}, 140506 (2007).



\bibitem{Nishino-CTMRG} T. Nishino and K. Okunishi, Corner transfer matrix renormalization group method, J. Phys. Soc. Jpn. \textbf{65}, 891 (1996).
\bibitem{Vidal-CTM} R. Orus, and G. Vidal,  Simulation of two-dimensional quantum systems on an infinite lattice
revisited: Corner transfer matrix for tensor contraction, Phys. Rev. B \textbf{80}, 094403 (2009).
\bibitem{Vari-CTMRG} X. F. Liu, Y. F. Fu, W. Q. Yu, J. F. Yu, Z. Y. Xie, Variational corner transfer matrix
renormalization group method for classical statistical models, Chin. Phys. Lett. \textbf{39}, 067502 (2022).
\bibitem{NTN2017} Z. Y. Xie, H. J. Liao, R. Z. Huang, H. D. Xie, J. Chen, Z. Y. Liu, and T. Xiang, Optimized
contraction scheme for tensor-network states, Phys. Rev. B \textbf{96}, 045128 (2017).
\bibitem{HM-MC} A. W. Sandvik, Computational studies of quantum spin systems, AIP Conf. Proc. \textbf{1297}, 135 (2010)
\bibitem{Corboz-SSM} P. Corboz, and F. Mila, Tensor network study of the Shastry-Sutherland model in zero magnetic
field, Phys. Rev. B \textbf{87}, 115144 (2013). 
\bibitem{BP1986} D. E. Rumelhart, G. E. Hinton, and R. J. Williams,  Learning representations by back propagating errors, Nature \textbf{323}, 533–536 (1986).
\bibitem{SRGAD2020} B. B. Chen, Y. Gao, Y. B. Guo, Y. Liu, H. H. Zhao, H. J. Liao, L. Wang, T. Xiang, W. Li, Z. Y. Xie, Automatic differentiation for second renormalization of tensor networks, Phys. Rev. B \textbf{101}, 220409(R) (2020).
\bibitem{LY-PRB2023} Y. Liu, Z. Y. Xie, H.-G. Luo, J. Zhao, Long-range spin-orbital order in the spin-orbital
$SU(2)\times SU(2)\times U(1)$ model, Phys. Rev. B \textbf{107}, L041106 (2023).
\bibitem{YJKao-CTMRG} C.-Y. Lee, B. Normand, and Y.-J. Kao, Gapless spin liquid in the kagome Heisenberg
antiferromagnet with Dzyaloshinskii-Moriya interactions, Phys. Rev. B \textbf{98}, 224414 (2018).




\bibitem{DNSheng-CTMRG} R. Haghshenas, S.-S. Gong, and D. N. Sheng, Single-layer tensor network study of the
Heisenberg model with chiral interactions on a kagome lattice, Phys. Rev. B \textbf{99}, 174423 (2019).
\bibitem{Vidal-TEBD} G. Vidal, Classical simulation of infinite-size quantum lattice systems in one spatial dimension, Phys. Rev. Lett. \textbf{98}, 070201 (2007).
\bibitem{Orus-TEBD} R. Orus, and G. Vidal, Infinite time-evolving block decimation algorithm beyond unitary
evolution, Phys. Rev. B \textbf{78}, 155117 (2008).
\bibitem{Explain} In principle, the dimension of each bond can be different. For example, in the case of Fig.~\ref{Fig:NTN}(c), all the bonds have dimension $D$ except the left and down bonds of tensor $Y$, which are of dimension $Dd$. In the main text, we assume all the bonds have the same dimension just for simplicity of description. In the meantime, since $d$ is almost always much smaller than $D$, we ignore the scaling of $d$ when the computational cost and the memory cost are considered in the main text.
\bibitem{Corboz-vf} P. Corboz, Variational optimization with infinite projected entangled-pair states, Phys. Rev. B \textbf{94}, 035133 (2016).
\bibitem{Corboz-SSMH} P. Corboz, and F. Mila, Crystals of bound states in the magnetization plateaus of the Shastry-Sutherland model, Phys. Rev. Lett. \textbf{112}, 147203 (2014).
\bibitem{WYL-PRL2024} W.-Y. Liu, X.-T. Zhang, Z. Wang, S.-S. Gong, W.-Q. Chen and Z.-C. Gu, Quantum criticality
with emergent symmetry in the extended Shastry-Sutherland model, Phys. Rev. Lett. \textbf{133}, 026502 (2024).
\bibitem{DQCP-Senthil} T. Senthil, A. Vishwanath, L. Balents, S. Sachdev, and M. P. A. Fisher, Deconfined quantum
critical points, Science \textbf{303}, 1490–1494 (2004).
\bibitem{LingWang-SSM} J. Yang, A. W. Sandvik, and L. Wang, Quantum criticality and spin liquid phase in the
Shastry-Sutherland model, Phys. Rev. B \textbf{105}, L060409 (2022).
\bibitem{Corboz-QSLSSM2025} P. Corboz, Y. Zhang, B. Ponsioen, and F. Mila, Quantum spin liquid phase in the Shastry-Sutherland model revealed by high-precision infinite projected entangled-pair states, arXiv:2502.14091.






\bibitem{SCBO-Science} Y. Cui, L. Liu, H. Lin, K.-H. Wu, W. Hong, X. Liu, C. Li, Z. Hu, N. Xi, S. Li,et al., Proximate deconfined quantum critical point in  SrCu$_2$(BO$_3$)$_2$, Science \textbf{380}, 1179 (2023).

\bibitem{NQS} L. L. Viteritti, R. Rende, A. Parola, S. Goldt, and F. Becca, Transformer wave function for two dimensional frustrated magnets: Emergence of a spin-liquid phase in the Shastry-Sutherland
model, Phys. Rev. B \textbf{111}, 134411 (2025).
\bibitem{ADPEPS} B. Ponsioen, F. F. Assaad, and P. Corboz, Automatic differentiation applied to excitations with
projected entangled pair states, SciPost Phys. \textbf{12}, 006 (2022).
\bibitem{Hasik-Sci2021} J. Hasik, D. Poilblanc, and F. Becca, Investigation of the Néel phase of the frustrated Heisenberg
antiferromagnet by differentiable symmetric tensor networks, SciPost Phys. \textbf{10}, 012 (2021).
\bibitem{LW-fixedMPS} H. Xie, J.-G. Liu, L. Wang, Automatic differentiation of dominant eigensolver and its
applications in quantum physics, Phys. Rev. B \textbf{101}, 245139 (2020).
\bibitem{SRG2010} H. H. Zhao, Z. Y. Xie, Q. N. Chen, Z. C. Wei, J. W. Cai, T. Xiang, Renormalization of
tensor-network states, Phys. Rev. B \textbf{81}, 174411 (2010).
\bibitem{RSVD} S. Morita, R. Igarashi, H.-H. Zhao, and N. Kawashima, Tensor renormalization group with
randomized singular value decomposition, Phys. Rev. E \textbf{97}, 033310 (2018).
\bibitem{Corboz-QR2025} Y. Zhang, Q. Yang, P. Corboz, Accelerating two-dimensional tensor network contractions
using QR-decompositions, arXiv:2505.00494.
\bibitem{Corboz-Z2} O. van Alphen, S. V. Kleijweg, J. Hasik, P. Corboz, Exploiting the Hermitian symmetry in
tensor network algorithms, Phys. Rev. B \textbf{111}, 045105 (2025).
\bibitem{Schmoll-PRB2025} J. Naumann, E. Lennart Weerda, J. Eisert, M. Rizzi, P. Schmoll, Variationally optimizing infinite
projected entangled-pair states at large bond dimensions: A split corner transfer matrix
renormalization group approach, Phys. Rev. B \textbf{111}, 235116 (2025).
\bibitem{ZhangXY-PRB2026} X.-Y. Zhang, Q. Y., P. Corboz, J. Haegeman, and W. Tang, Accelerating two-dimensional
tensor network optimization by preconditioning, Phys. Rev. B \textbf{113}, 125111 (2026).
\bibitem{ZGM-PRB2025} X.-Z. Xu, T.-Y. Lin, G.-M. Zhang, Efficient optimization of variational tensor-network
approach to three-dimensional statistical systems, Phys. Rev. B \textbf{112}, 134403 (2025).
\bibitem{YLP-PRB2023} L.-P. Yang, Y. F. Fu, Z. Y. Xie, T. Xiang, Efficient calculation of three-dimensional tensor
networks, Phys. Rev. B \textbf{107}, 165127 (2023).

\end{thebibliography}
\end{document}